\documentclass[modern]{aastex61}

\usepackage{color}

\newcommand{\GG}[1]{}

\usepackage{float}
\usepackage{placeins}

\submitjournal{ApJ}

\begin{document}

\title{The Great Markarian 421 Flare of February 2010: Multiwavelength variability and correlation studies}


\correspondingauthor{Lucy Fortson}
\email{lfortson@umn.edu}

\author{A.~U.~Abeysekara}\affiliation{Department of Physics and Astronomy, University of Utah, Salt Lake City, UT 84112, USA}
\author{W.~Benbow}\affiliation{Center for Astrophysics | Harvard \& Smithsonian, Fred Lawrence Whipple Observatory, Amado, AZ 85645, USA}
\author{R.~Bird}\affiliation{Department of Physics and Astronomy, University of California, Los Angeles, CA 90095, USA}
\author{A.~Brill}\affiliation{Physics Department, Columbia University, New York, NY 10027, USA}
\author{R.~Brose}\affiliation{Institute of Physics and Astronomy, University of Potsdam, 14476 Potsdam-Golm, Germany and DESY, Platanenallee 6, 15738 Zeuthen, Germany}
\author{M.~Buchovecky}\affiliation{Department of Physics and Astronomy, University of California, Los Angeles, CA 90095, USA}
\author{J.~H.~Buckley}\affiliation{Department of Physics, Washington University, St. Louis, MO 63130, USA}
\author{J.~L.~Christiansen}\affiliation{Physics Department, California Polytechnic State University, San Luis Obispo, CA 94307, USA}
\author{A.~J.~Chromey}\affiliation{Department of Physics and Astronomy, Iowa State University, Ames, IA 50011, USA}
\author{M.~K.~Daniel}\affiliation{Center for Astrophysics | Harvard \& Smithsonian, Fred Lawrence Whipple Observatory, Amado, AZ 85645, USA}
\author{J.~Dumm}\affiliation{School of Physics and Astronomy, University of Minnesota, Minneapolis, MN 55455, USA}
\author{A.~Falcone}\affiliation{Department of Astronomy and Astrophysics, 525 Davey Lab, Pennsylvania State University, University Park, PA 16802, USA}
\author{Q.~Feng}\affiliation{Physics Department, Columbia University, New York, NY 10027, USA}
\author{J.~P.~Finley}\affiliation{Department of Physics and Astronomy, Purdue University, West Lafayette, IN 47907, USA}
\author{L.~Fortson}\affiliation{School of Physics and Astronomy, University of Minnesota, Minneapolis, MN 55455, USA}
\author{A.~Furniss}\affiliation{Department of Physics, California State University - East Bay, Hayward, CA 94542, USA}
\author{N.~Galante}\affiliation{Center for Astrophysics | Harvard \& Smithsonian, Fred Lawrence Whipple Observatory, Amado, AZ 85645, USA}
\author{A.~Gent}\affiliation{School of Physics and Center for Relativistic Astrophysics, Georgia Institute of Technology, 837 State Street NW, Atlanta, GA 30332-0430}
\author{G.~H.~Gillanders}\affiliation{School of Physics, National University of Ireland Galway, University Road, Galway, Ireland}
\author{C.~Giuri}\affiliation{DESY, Platanenallee 6, 15738 Zeuthen, Germany}
\author{O.~Gueta}\affiliation{DESY, Platanenallee 6, 15738 Zeuthen, Germany}
\author{T.~Hassan}\affiliation{DESY, Platanenallee 6, 15738 Zeuthen, Germany}
\author{O.~Hervet}\affiliation{Santa Cruz Institute for Particle Physics and Department of Physics, University of California, Santa Cruz, CA 95064, USA}
\author{J.~Holder}\affiliation{Department of Physics and Astronomy and the Bartol Research Institute, University of Delaware, Newark, DE 19716, USA}
\author{G.~Hughes}\affiliation{Center for Astrophysics | Harvard \& Smithsonian, Fred Lawrence Whipple Observatory, Amado, AZ 85645, USA}
\author{T.~B.~Humensky}\affiliation{Physics Department, Columbia University, New York, NY 10027, USA}
\author{C.~A.~Johnson}\affiliation{Santa Cruz Institute for Particle Physics and Department of Physics, University of California, Santa Cruz, CA 95064, USA}
\author{P.~Kaaret}\affiliation{Department of Physics and Astronomy, University of Iowa, Van Allen Hall, Iowa City, IA 52242, USA}
\author{P.~Kar}\affiliation{Department of Physics and Astronomy, University of Utah, Salt Lake City, UT 84112, USA}
\author{N.~Kelley-Hoskins}\affiliation{DESY, Platanenallee 6, 15738 Zeuthen, Germany}
\author{M.~Kertzman}\affiliation{Department of Physics and Astronomy, DePauw University, Greencastle, IN 46135-0037, USA}
\author{D.~Kieda}\affiliation{Department of Physics and Astronomy, University of Utah, Salt Lake City, UT 84112, USA}
\author{M.~Krause}\affiliation{DESY, Platanenallee 6, 15738 Zeuthen, Germany}
\author{F.~Krennrich}\affiliation{Department of Physics and Astronomy, Iowa State University, Ames, IA 50011, USA}
\author{S.~Kumar}\affiliation{Physics Department, McGill University, Montreal, QC H3A 2T8, Canada}
\author{M.~J.~Lang}\affiliation{School of Physics, National University of Ireland Galway, University Road, Galway, Ireland}
\author{P.~Moriarty}\affiliation{School of Physics, National University of Ireland Galway, University Road, Galway, Ireland}
\author{R.~Mukherjee}\affiliation{Department of Physics and Astronomy, Barnard College, Columbia University, NY 10027, USA}
\author{T.~Nelson}\affiliation{School of Physics and Astronomy, University of Minnesota, Minneapolis, MN 55455, USA}
\author{D.~Nieto}\affiliation{Physics Department, Columbia University, New York, NY 10027, USA}
\author{M.~Nievas-Rosillo}\affiliation{DESY, Platanenallee 6, 15738 Zeuthen, Germany}
\author{S.~O'Brien}\affiliation{School of Physics, University College Dublin, Belfield, Dublin 4, Ireland}
\author{R.~A.~Ong}\affiliation{Department of Physics and Astronomy, University of California, Los Angeles, CA 90095, USA}
\author{A.~N.~Otte}\affiliation{School of Physics and Center for Relativistic Astrophysics, Georgia Institute of Technology, 837 State Street NW, Atlanta, GA 30332-0430}
\author{N.~Park}\affiliation{WIPAC and Department of Physics, University of Wisconsin-Madison, Madison WI, USA}
\author{A.~Petrashyk}\affiliation{Physics Department, Columbia University, New York, NY 10027, USA}
\author{A.~Pichel}\affiliation{Instituto de Astronomía y Física del Espacio (IAFE, CONICET-UBA), CC 67 - Suc. 28, (C1428ZAA) Ciudad Autónoma de Buenos Aires, Argentina}
\author{M.~Pohl}\affiliation{Institute of Physics and Astronomy, University of Potsdam, 14476 Potsdam-Golm, Germany and DESY, Platanenallee 6, 15738 Zeuthen, Germany}
\author{R.~R.~Prado}\affiliation{DESY, Platanenallee 6, 15738 Zeuthen, Germany}
\author{E.~Pueschel}\affiliation{DESY, Platanenallee 6, 15738 Zeuthen, Germany}
\author{J.~Quinn}\affiliation{School of Physics, University College Dublin, Belfield, Dublin 4, Ireland}
\author{K.~Ragan}\affiliation{Physics Department, McGill University, Montreal, QC H3A 2T8, Canada}
\author{P.~T.~Reynolds}\affiliation{Department of Physical Sciences, Cork Institute of Technology, Bishopstown, Cork, Ireland}
\author{G.~T.~Richards}\affiliation{Department of Physics and Astronomy and the Bartol Research Institute, University of Delaware, Newark, DE 19716, USA}
\author{E.~Roache}\affiliation{Center for Astrophysics | Harvard \& Smithsonian, Fred Lawrence Whipple Observatory, Amado, AZ 85645, USA}
\author{A.~C.~Rovero}\affiliation{Instituto de Astronomía y Física del Espacio (IAFE, CONICET-UBA), CC 67 - Suc. 28, (C1428ZAA) Ciudad Autónoma de Buenos Aires, Argentina}
\author{C.~Rulten}\affiliation{School of Physics and Astronomy, University of Minnesota, Minneapolis, MN 55455, USA}
\author{I.~Sadeh}\affiliation{DESY, Platanenallee 6, 15738 Zeuthen, Germany}
\author{M.~Santander}\affiliation{Department of Physics and Astronomy, University of Alabama, Tuscaloosa, AL 35487, USA}
\author{G.~H.~Sembroski}\affiliation{Department of Physics and Astronomy, Purdue University, West Lafayette, IN 47907, USA}
\author{K.~Shahinyan}\affiliation{School of Physics and Astronomy, University of Minnesota, Minneapolis, MN 55455, USA}
\author{B.~Stevenson}\affiliation{Department of Physics and Astronomy, University of California, Los Angeles, CA 90095, USA}
\author{I.~Sushch}\affiliation{Institute of Physics and Astronomy, University of Potsdam, 14476 Potsdam-Golm, Germany}
\author{J.~Tyler}\affiliation{Physics Department, McGill University, Montreal, QC H3A 2T8, Canada}
\author{V.~V.~Vassiliev}\affiliation{Department of Physics and Astronomy, University of California, Los Angeles, CA 90095, USA}
\author{S.~P.~Wakely}\affiliation{Enrico Fermi Institute, University of Chicago, Chicago, IL 60637, USA}
\author{A.~Weinstein}\affiliation{Department of Physics and Astronomy, Iowa State University, Ames, IA 50011, USA}
\author{R.~M.~Wells}\affiliation{Department of Physics and Astronomy, Iowa State University, Ames, IA 50011, USA}
\author{P.~Wilcox}\affiliation{Department of Physics and Astronomy, University of Iowa, Van Allen Hall, Iowa City, IA 52242, USA}
\author{A.~Wilhelm}\affiliation{Institute of Physics and Astronomy, University of Potsdam, 14476 Potsdam-Golm, Germany and DESY, Platanenallee 6, 15738 Zeuthen, Germany}
\author{D.~A.~Williams}\affiliation{Santa Cruz Institute for Particle Physics and Department of Physics, University of California, Santa Cruz, CA 95064, USA}
\author{B.~Zitzer}\affiliation{Physics Department, McGill University, Montreal, QC H3A 2T8, Canada}

\collaboration{(VERITAS Collaboration)}

\author{V.~A.~Acciari}
\affiliation{Inst. de Astrof\'isica de Canarias, E-38200 La Laguna, and Universidad de La Laguna, Dpto. Astrof\'isica, E-38206 La Laguna, Tenerife, Spain}

\author{S.~Ansoldi}
\affiliation{Universit\`a di Udine, and INFN Trieste, I-33100 Udine, Italy}
\affiliation{Japanese MAGIC Consortium: ICRR, The University of Tokyo, 277-8582 Chiba, Japan; Department of Physics, Kyoto University, 606-8502 Kyoto, Japan; Tokai University, 259-1292 Kanagawa, Japan; RIKEN, 351-0198 Saitama, Japan}

\author{L.~A.~Antonelli}
\affiliation{National Institute for Astrophysics (INAF), I-00136 Rome, Italy}

\author{A.~Arbet Engels}
\affiliation{ETH Zurich, CH-8093 Zurich, Switzerland}

\author{D.~Baack}
\affiliation{Technische Universit\"at Dortmund, D-44221 Dortmund, Germany}

\author{A.~Babi\'c}
\affiliation{Croatian MAGIC Consortium: University of Rijeka, 51000 Rijeka; University of Split - FESB, 21000 Split; University of Zagreb - FER, 10000 Zagreb; University of Osijek, 31000 Osijek; Rudjer Boskovic Institute, 10000 Zagreb, Croatia}

\author{B.~Banerjee}
\affiliation{Saha Institute of Nuclear Physics, HBNI, 1/AF Bidhannagar, Salt Lake, Sector-1, Kolkata 700064, India}

\author{U.~Barres de Almeida}
\affiliation{Centro Brasileiro de Pesquisas F\'isicas (CBPF), 22290-180 URCA, Rio de Janeiro (RJ), Brasil}

\author{J.~A.~Barrio}
\affiliation{Unidad de Part\'iculas y Cosmolog\'ia (UPARCOS), Universidad Complutense, E-28040 Madrid, Spain}

\author{J.~Becerra Gonz\'alez}
\affiliation{Inst. de Astrof\'isica de Canarias, E-38200 La Laguna, and Universidad de La Laguna, Dpto. Astrof\'isica, E-38206 La Laguna, Tenerife, Spain}

% ...

\collaboration{(MAGIC Collaboration)}

\author{H. D. Aller}
\affiliation{Department of Astronomy, University of Michigan, Ann Arbor, MI 48109-1107, USA}

\author{M. F. Aller}
\affiliation{Department of Astronomy, University of Michigan, Ann Arbor, MI 48109-1107, USA}

\author{M.~T.~Carini}
\affiliation{Department of Physics and Astronomy, Western Kentucky University, 1906 College Heights Boulevard \#11077, Bowling Green, KY 42101, USA}

\author{D.~Horan}
\affiliation{Laboratoire Leprince-Ringuet, \'Ecole Polytechnique, CNRS/IN2P3, F-91128 Palaiseau, France}

\author{B. Jordan}
\affiliation{School of Cosmic Physics, Dublin Institute For Advanced Studies, Ireland}

\author{Svetlana G. Jorstad}
\affiliation{Institute for Astrophysical Research, Boston University, 725 Commonwealth Avenue, Boston, MA 02215}
\affiliation{Astronomical Institute, St. Petersburg State University, Universitetskij Pr. 28, Petrodvorets, 198504 St. Petersburg, Russia}

\author{O.M. Kurtanidze}
\affiliation{Abastumani Observatory, Mt. Kanobili, 0301 Abastumani, Georgia}
\affiliation{Engelhardt Astronomical Observatory, Kazan Federal University,   Tatarstan, Russia}
\affiliation{Center for Astrophysics, Guangzhou University, Guangzhou 510006, China}

\author{S.O. Kurtanidze}
\affiliation{Abastumani Observatory, Mt. Kanobili, 0301 Abastumani, Georgia}

\author{A. L\"ahteenm\"aki}
\affiliation{Aalto University Mets\"ahovi Radio Observatory, Mets\"ahovintie 114,
FI-02540 Kylm\"al\"a, Finland}
\affiliation{Aalto University Department of Electronics and Nanoengineering, P.O. BOX 15500, FI-00076 AALTO, Finland}

\author{V.~M.~Larionov}
\affiliation{Astronomical Institute, St. Petersburg State University, Universitetskij Pr. 28, Petrodvorets, 198504 St. Petersburg, Russia}
\affiliation{Pulkovo Observatory, St. Petersburg, Russia}
\author{E.~G.~Larionova}  
\affiliation{Astronomical Institute, St. Petersburg State University, Universitetskij Pr. 28, Petrodvorets, 198504 St. Petersburg, Russia}

\author{G.~Madejski}
\affiliation{W.~W.~Hansen Experimental Physics Laboratory, Kavli Institute for Particle Astrophysics and Cosmology, Department of Physics and SLAC National Accelerator Laboratory, Stanford University, Stanford, CA 94305, USA}

\author{Alan P. Marscher} 
\affiliation{Institute for Astrophysical Research, Boston University, 725 Commonwealth Avenue, Boston, MA 02215}

\author{W.~Max-Moerbeck}
\affiliation{Universidad de Chile, Departamento de Astronomía, Camino El Observatorio 1515, Las Condes, Santiago, Chile}

\author{J. Ward Moody}
\affiliation{Department of Physics and Astronomy, Brigham Young University, Provo, Utah 84602, USA}

\author{D.~A.~Morozova}  
\affiliation{Astronomical Institute, St. Petersburg State University, Universitetskij Pr. 28, Petrodvorets, 198504 St. Petersburg, Russia}

\author{M.G. Nikolashvili}
\affiliation{Abastumani Observatory, Mt. Kanobili, 0301 Abastumani, Georgia}

\author{C.~M.~Raiteri}
\affiliation{INAF, Osservatorio Astrofisico di Torino, I-10025 Pino Torinese (TO), Italy}

\author{A. C. S. Readhead}
\affiliation{Cahill Centre for Astronomy and Astrophysics, California Institute of Technology, Pasadena, CA 91125, USA}

\author{J. L. Richards}
\affiliation{Department of Physics and Astronomy, Purdue University, West Lafayette, IN 47907, USA}

\author{Alberto C. Sadun}
\affiliation{Department of Physics, University of Colorado Denver, Denver, Colorado, CO 80217-3364, USA}

% GRT
\author{T. Sakamoto}
\affiliation{Department of Physics and Mathematics, College of Science and Engineering, Aoyama Gakuin University, 5-10-1 Fuchinobe,Chuoku, Sagamihara-shi Kanagawa 252-5258, Japan}

\author{L.A. Sigua}
\affiliation{Abastumani Observatory, Mt. Kanobili, 0301 Abastumani, Georgia}

% Steward
\author{P.~S.~Smith}
\affiliation{Steward Observatory, University of Arizona, Tucson, AZ 85721 USA}

\author{H.~Talvikki}
\affiliation{Tuorla Observatory, Department of Physics and Astronomy, University of Turku, 20014 Turku, Finland}

\author{J. Tammi}
\affiliation{Aalto University Mets\"ahovi Radio Observatory, Mets\"ahovintie 114,
FI-02540 Kylm\"al\"a, Finland}

\author{M. Tornikoski}
\affiliation{Aalto University Mets\"ahovi Radio Observatory, Mets\"ahovintie 114,
FI-02540 Kylm\"al\"a, Finland}
 
\author{I.~S.~Troitsky}  
\affiliation{Astronomical Institute, St. Petersburg State University, Universitetskij Pr. 28, Petrodvorets, 198504 St. Petersburg, Russia}

%% Optical %%
%GASP-WEBT
\author{M.~Villata}
\affiliation{INAF, Osservatorio Astrofisico di Torino, I-10025 Pino Torinese (TO), Italy}

\collaboration{(Multiwavelength Partners)}



\begin{abstract}

We report on variability and correlation studies using multiwavelength observations of the blazar Mrk\,421 during the month of February, 2010 when an extraordinary flare reaching a level of $\sim$27~Crab Units above 1~TeV was measured in very-high-energy (VHE) $\gamma$-rays with the VERITAS observatory. This is the highest flux state for Mrk\,421 ever observed in VHE $\gamma$-rays. Data are analyzed from a coordinated campaign across multiple instruments including VHE $\gamma$-ray (VERITAS, MAGIC), high-energy (HE) $\gamma$-ray (\textit{Fermi}-LAT), X-ray (\textit{Swift}, \textit{RXTE}, MAXI), optical (including the GASP-WEBT collaboration and polarization data) and radio (Mets\"ahovi, OVRO, UMRAO).
Light curves are produced spanning multiple days before and after the peak of the VHE flare, including over several flare `decline' epochs.  
The main flare statistics allow 2-minute time bins to be constructed in both the VHE and optical bands enabling a cross-correlation analysis that shows evidence for an optical lag of $\sim$25--55 minutes, the first time-lagged correlation between these bands reported on such short timescales. Limits on the Doppler factor ($\delta \gtrsim 33$) and the size of the emission region ($ \delta^{-1}R_B \lesssim 3.8\times 10^{13}\,\,\mbox{cm}$) are obtained from the fast variability observed by VERITAS during the main flare. Analysis of 10-minute-binned VHE and  X-ray data over the decline epochs shows an extraordinary range of behavior in the flux-flux relationship: from linear to quadratic to lack of correlation to anti-correlation. Taken together, these detailed observations of an unprecedented flare seen in Mrk\,421 are difficult to explain by the classic single-zone synchrotron self-Compton model.

\end{abstract}


\section{Introduction} \label{introduction}

Blazars are a sub-class of radio-loud active galactic nuclei (AGN) with jets of relativistic material beamed nearly along the line-of-sight \citep{Blandford1978,Urry1995} whose non-thermal radiation is observed across the entire spectrum, from radio to $\gamma$-rays. Due to Doppler beaming, the bolometric luminosity of blazars can be dominated by very-high-energy (VHE; $>100$~GeV) $\gamma$-rays. At a redshift of \emph{z}=0.031, Mrk\,421 is the closest known BL Lac object \citep{devau91} and is the first extragalactic object to be detected in VHE $\gamma$-rays \citep{Punch92}. Blazars now comprise the majority source class of VHE extragalactic $\gamma$-ray emitters \citep{Wakely2008} and while there is much we have learned from multiwavelength data taken over the past forty years on Mrk\,421 and other  blazars, there remain many unanswered questions. Indeed, there is still no general agreement on the particle acceleration mechanism within the jet or the location of $\gamma$-ray emission zone(s)~\citep[e.g.,][]{Boettcher2019}.  Nonetheless,  progress can be made 
 through dedicated campaigns organized simultaneously across as many wavebands as possible~\citep[e.g.,][]{Aleksic2015mid,Furniss2015,Ahnen2018}. 
 
 The spectral energy distribution (SED) of blazars is characterized by a double peak where the lower peak is due to synchrotron radiation while the higher peak is generally thought to arise from inverse-Compton (IC) upscattering of lower-energy photons off the population of accelerating electrons in the jet \citep{Jones74}. Hadronic models \citep{Aharonian2000,Mucke2001,Mannheim1993,Dimitrakoudis2014} or even lepto-hadronic models \citep{Cerruti2015}, may also be responsible for the second SED peak. The Synchrotron-Self Compton (SSC) model posits that the seed photons for the IC process are the synchrotron photons from the accelerating electrons \citep[e.g.,][]{Ghisellini1998}. Observationally, blazars are classified by the peak frequency of their synchrotron emission;  with ${\nu}_{s}=10^{18.9}$~Hz, Mrk\,421 is deemed a high-frequency peaked BL Lac (HBL) \citep{Nieppola06}. 
  
Blazars exhibit complex temporal structures with strong variability across the spectrum from radio to $\gamma$-rays (e.g.~\cite{Romero2017} and references therein). Blazar light curves are typically aperiodic with power-law Power Spectral Density (PSD) distributions indicative of stochastic processes \citep{Finke2015}.  Multi-band blazar light curves can be punctuated by dramatic flares on timescales from minutes to days where inter-band correlation is often observed  \citep{Acciari2011, Abramowski2012, Ahnen2018}. 

Studying the time-varying characteristics of a source through  multiwavelength campaigns can test model predictions on what governs the $\gamma$-ray emission and its location within the jet. The standard homogeneous single-zone SSC model of blazar emission employs a single population of electrons that is accelerated in a compact region $< 1$ pc from the central engine (the central black hole driving the jet). The accelerated electrons cool through the emission of synchrotron radiation, then potentially through IC scattering and/or escape out of the accelerating ``blob".
The spatial scale of the emission region can be set by the variability detected in the VHE-band observations. Competition between cooling, acceleration and dynamical timescales that characterize the system can lead to several potential observables including asymmetries in flare profiles and ``soft'' or ``hard'' lags (and accompanying clockwise or counter-clockwise hysteresis loops) as described in e.g.\ \cite{kirk98} or \cite{LiKusunose2000}. 

Much of the previous work with Mrk\,421 as well as the ever-growing population of blazars detected by VHE instruments indicates that most SEDs of HBLs can be described by a single-zone SSC model
 \citep{Beilicke11,Abeysekara2017,Ahnen2018}. 
As tracers of the same underlying electron population, hard X-rays typically probe the falling edge of the synchrotron peak while VHE $\gamma$-rays probe the falling edge of the IC peak in an HBL with the expectation that these bands will show highly correlated fast variability.  
However, orphan flares, such as the 2002 VHE flare observed in 1ES 1959+650 \citep{Krawczynski2004} without a corresponding X-ray flare, provide evidence  that one-zone SSC models are too simplistic. The remarkable VHE flare in PKS 2155-304 seen by the High Energy Stereoscopic System (H.E.S.S.) in late July, 2006 suggests a need for two emission zones to explain the data \citep{Aharonian2009a}. Several recent campaigns on Mrk\,421 and Mrk\,501 also indicate  a preference for a multi-component scenario~\citep{Aleksic2015March2010,Ahnen2017}.

Fast flaring events provide another test of the SSC model.
Several blazars have been observed to emit VHE flares that vary on timescales of 5-20 minutes \citep{Albert2007, Aharonian2009a}. Mrk\,421 itself has a history of fast flares including those reported in \cite{Gaidos1996,Blazejowski2005, Fossati2008, Beilicke11}. These $\sim$minute timescale flares pose serious issues for single-zone blazar models as the implied high bulk Lorentz factors required are in tension with the radio observations of these sources \citep{Boettcher2013,Piner2018}.
Moreover, the shock-in-jet model suggested to explain knots of material traveling along the jet in radio observations is found to be incompatible with the highly-compact emission regions implied by fast flaring episodes detected in blazars \citep{Romero2017}. Indeed, since the majority of blazars are detected during flaring episodes, the erroneous interpretation could be made that a single-zone SSC scenario is responsible for the generic form of the object's SED. In fact, there may be more than one emitting region at any given time, with one region accounting for ``quiescent" or ``envelope'' behavior while another region or process may be responsible for a detected flare triggered by a localized event (e.g.\,magnetic reconnection~\citep{petropoulou16}). 
    
Given the sometimes surprising and dynamical nature of blazars, efforts to coordinate multiwavelength campaigns continue to be important. The results from each campaign provide further clues for modelers to incorporate.
For example, highly correlated rapid variability observed between the VHE and optical bands such as described in this work, has not been reported before, and has not been accounted for in modeling.
The observed (or lack of observed) correlated activity between specific bands can discriminate between possible emission mechanisms; and stringent constraints on the sizes and locations of $\gamma$-ray emission regions can be set by the flux and spectral variability patterns of blazars~\citep{Boettcher2012}.

In this paper, we apply timing analysis techniques, including variability and correlation studies, to the  extraordinary Mrk\,421 flare recorded in February, 2010 by the VERITAS observatory and many multiwavelength partners. 
During 2009-2010, Mrk\,421 was the object of an intense multiwavelength campaign organized by the \textit{Fermi} Large Area Telescope (\textit{Fermi}-LAT) collaboration and involving the ground-based imaging air-Cherenkov telescopes (IACTs) (H.E.S.S., MAGIC and VERITAS) as well as \textit{RXTE} and \textit{Swift} satellites in X-ray, \textit{Swift} UVOT ultraviolet, and numerous ground-based optical
and radio telescopes. Several smaller flares were observed throughout the campaign including one in March, 2010 described in \citep{Aleksic2015March2010}.
Here we report on the multiwavelength dataset covering the period 1 February - 1 March, 2010 UT (MJD 55228 - 55256) with a focus on the giant VHE flare on 17 February, 2010 UT (MJD 55244). 
We note that several other instruments have observed the same flare including MAXI \citep{Isobe2010}, H.E.S.S.  \citep{Tluczy10},  HAGAR \citep{Shukla2012} and TACTIC \citep{Singh2015}. 

This paper is organized as follows:~in Section~\S\ref{datasets} we describe the multiwavelength datasets including the respective methods for analyzing the data presented. In Section~\S\ref{tevresults} we focus on the results from the night of the exceptional flare including the variability analysis of VERITAS data as well as results from optical-VHE correlation studies. The results of further multiwavelength studies over the full February 2010 dataset are presented in Section~\S\ref{mwlresults}, including multiwavelength variability studies as well as VHE-X-ray and HE-X-ray correlation analyses. We conclude with an overall discussion of the results in 
Section~\S\ref{conclusions}.

%%%%%%%%%%%%%%%%%%%%%%%%%%%%%%%%%%%%%%%%%%%%%%%%%%%%%%%%%%%%%%%%%%%%%%
%%%%%%%%%%%%%%%%%%%%%%%%%%%%%%%%%%%%%%%%%%%%%%%%%%%%%%%%%%%%%%%%%%%%%%
%%%%%%%%%%%%%%%%%%%%%%%%%%%%%%%%%%%%%%%%%%%%%%%%%%%%%%%%%%%%%%%%%%%%%%
\section{Datasets and Data Reduction} \label{datasets}

\begin{figure}[h]
\epsscale{1.25}
\plotone{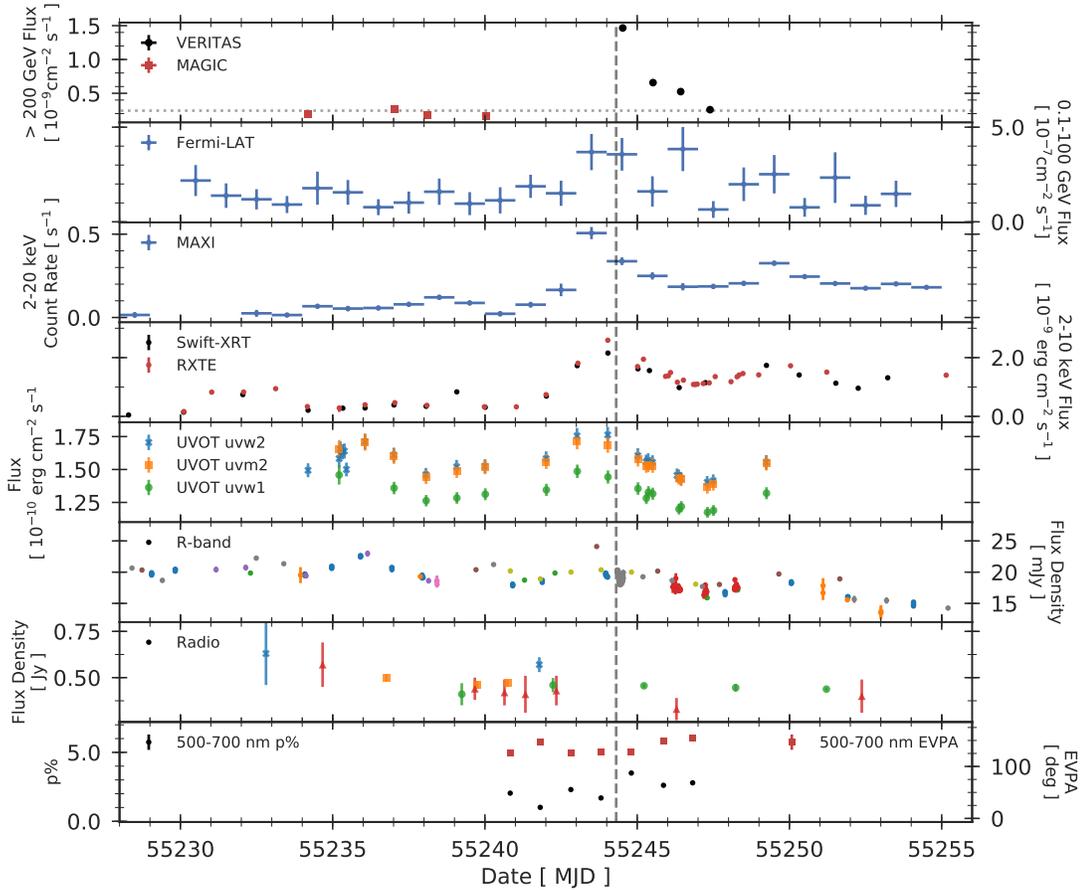}
\caption{Light curves for multi-band observations during the February, 2010 portion of the \textit{Fermi}-LAT led campaign. From top to bottom: VHE (VERITAS, MAGIC), HE (\textit{Fermi}-LAT), hard X-ray (MAXI), X-ray (\textit{RXTE}, \textit{Swift}-XRT), UV (\textit{Swift}-UVOT), optical (Abastumani (blue), CRAO (orange), GRT (green), GalaxyView (red), KVA (purple), NewMexicoSkies (brown), Perkins (pink), RCT (gray), and Steward (tan)), and radio (UMRAO 8~GHz (blue stars) and 14~GHz (orange squares), OVRO 15~GHz (green circles) and Mets{\"a}hovi 37~GHz (red triangles)), including optical polarization observations (Steward Observatory). The light curves are binned by individual observation except for VHE, HE, and MAXI which are daily binned. The time of the giant VHE flare is denoted by the dashed vertical line. The dotted horizontal line on the VHE (top panel) denotes 1 Crab Unit based on \cite{Aharonian2006c}. Note that error bars are not visible for several bands due to high statistics.}  
\label{fig:mwlLC}
\end{figure}

The multiwavelength light curves covering radio-to-VHE observations around the time of the Mrk\,421 February, 2010 flare are shown in Figure~\ref{fig:mwlLC}. While the light curves in Figure~\ref{fig:mwlLC} are meant as an overview of available observations, they demonstrate the full breadth of the campaign and show the progression of the flare; more detailed light curves in the various wavebands are considered later in the paper. We summarize the available datasets and present details of the instruments in the following subsections. The data for light curves used throughout this paper are available through the on-line version of the article.

%%%% FIGURE 1 %%%%%%%%

%%%%%%%%%%%%%%%%%%%%%%%%%%%%%%%%%%%%%%%%%%%%%%%%%%%%%%%%%%%%%%%%%%%%%%
\subsection{ VHE $\gamma$-ray observations}
\label{DataVHE}
The VHE $\gamma$-ray data comprise both MAGIC and VERITAS observations.  Starting with the MAGIC observatory, %as part of an extensive MWL campaign, 
data were taken on Mrk\,421 between MJD 55234 and 55240 (7 February and 13 February, 2010) with
bad weather preventing further observations.
Upon alert from the campaign that the X-ray state was quite high and variable, VERITAS picked up the observations between MJD 55244 and 55247 (17 February and 20 February, 2010). 
As soon as VERITAS began taking Mrk\,421 data on MJD 55244, VERITAS observed a remarkable flare in progress with a peak flux $\sim 15$ Crab Units (CU) above 200~GeV~\citep[CU based on][]{Aharonian2006c}. 
Over the next two days, VERITAS observed the flux decrease to the average over the 2009-2010 season, $\sim1$~CU, which was itself an elevated state \citep [see][]{Beilicke11}. 
The MAGIC and VERITAS combined
VHE $\gamma$-ray light curve for Mrk\,421 is shown  in the top panel of Figure \ref{fig:mwlLC} in daily bins for data taken between MJD 55234 and 55247.
The VHE data with considerably finer binning are described below in Sections~\S\ref{VHESpectra}, \S\ref{OpticalCorrelation} and~\S\ref{VHE-xray-Correlation}.

%%%%%%%%%%%%%%%%%%%%%%%%%%%%%%%%%%%%%%%%%%%%%%%%%%%%%%% 
\paragraph{VERITAS.}

The VERITAS  array \citep{Holder06, Acciari08} is located at 1300~m above sea level in Arizona at the Fred Lawrence Whipple Observatory 
(31$^\circ$ 40$'$ N, 110$^{\circ}$ 57$'$ W) and comprises four Davies-Cotton-design telescopes
each with a 12-m diameter primary reflector. 
During the observations presented here, 
VERITAS was sensitive to $\gamma$-rays between 100~GeV and tens of TeV with an energy resolution better than $\sim$20\%, and an integral flux sensitivity that would have allowed a point source detection with a 1\% Crab Nebula flux in less than 30 hours.

A total of 17 hours 21 minutes of Mrk\,421 data were taken by VERITAS over the month of February of which 16 hours 44 minutes were data taken in good weather conditions. A total of 5 hours 12 minutes of data were collected on the night of the giant flare (MJD 55244), with observations starting at $83^{\circ}$ elevation and ending with an elevation of $40^{\circ}$, giving rise to a higher-energy threshold for later observations. Three of the four telescopes were operational during this time (on MJD 55244). All four telescopes participated in the rest of the February data and all observations were made in wobble mode \citep{Fomin1994}. 
Data were analyzed and cross-checked with the two standard VERITAS analysis packages~\citep{Cogan2008,Daniel2008}.

\paragraph{MAGIC.}

The Major Atmospheric Gamma-ray Imaging Cherenkov (MAGIC) telescope system consists of two telescopes each with a 17-m diameter mirror dish, 
located at 2200 m above sea level at the Roque de los Muchachos, on the Canary Island of La Palma ( 28$^{\circ}$ 46$'$ N, 18$^{\circ}$ 53$'$ W) \citep{MAGICPerformancePaper2008, MAGICperf12,Aleksic2016}. 

The MAGIC data for Mrk\,421 in February, 2010 comprise a total of 2 hours over four separate observations.
The data were taken in wobble mode at an elevation above 60$^{\circ}$ to achieve the lowest-possible energy threshold. 
These data were analyzed following the standard procedure \cite{MAGICperf12} with the MAGIC Analysis and Reconstruction Software \citep[MARS][]{Moralejo2009}. 

%%%%%%%%%%%%%%%%%%%%%%%%%%%%%%%%%%%%%%%%%%%%%%%%%%%%%%%
\subsection{HE $\gamma$-ray observations}

The Large Area Telescope (LAT) aboard the \textit{Fermi} Gamma-Ray Space Telescope (FGST) 
is a pair-conversion detector sensitive to $\gamma$-rays between 20 MeV
and 300 GeV. FGST typically operates in survey mode 
such that Mrk\,421 is observed once every $\sim3$ hours~\citep{Atwood2009}.

Events belonging to the \textit{Source} data class with energies between 100~MeV and 100~GeV were selected and analyzed using the \texttt{P8R2\_SOURCE\_V6} instrument response functions and \texttt{v10r0p5} of the \textit{Fermi} \texttt{ScienceTools}.\footnote{\url{http://fermi.gsfc.nasa.gov/ssc/data/analysis/documentation/Cicerone/}
   } 
In order to avoid contamination from Earth limb photons, a zenith angle cut of $<90^\circ$ was applied.  The analysis considered data from MJD 55230 to 55255, which is the 25-day period centered on the peak of the TeV flare as detected by VERITAS.  
   
The full dataset was analyzed using a binned likelihood analysis. The likelihood model included all \textit{Fermi}-LAT sources from the third \textit{Fermi} catalog \citep{3fgl} located within a 15$^\circ$ region-of-interest (RoI) centered on Mrk\,421, as well as the isotropic and Galactic diffuse emission. For the full dataset, Mrk\,421 was fitted with a power-law model, with both the flux normalization and photon index being left as free parameters in the likelihood fit. All spectral parameters were fixed for sources located at $>7^{\circ}$, while the normalization parameter was fitted for sources between $3^{\circ}$ and $7^{\circ}$, \added{and all parameters were fitted for sources $<3^{\circ}$ from the RoI center.}
   
The optimized RoI model from the full dataset was used to calculate the Mrk\,421 daily-binned light curve. The likelihood analysis was repeated for each time bin to obtain the daily flux points. Only the normalization parameter for Mrk\,421 was fitted, while all other RoI model parameters were kept fixed. The resulting Mrk\,421 light curve with daily binning is shown in the second panel from the top in Figure \ref{fig:mwlLC}. Fitting the index parameter of the Mrk\,421 model along with the normalization parameter has an insignificant impact on the result. 
   
%%%%%%%%%%%%%%%%%%%%%%%%%%%%%%%%%%%%%%%%%%%%%%%%%%%%%%%%%%%%%%%%%%%%%%
\subsection{X-ray observations}
\label{DataXRay}
We obtained X-ray data over the period of interest from three different observatories:\, MAXI, the Rossi X-ray Timing Experiment (\textit{RXTE}) and \textit{Swift}.  Just prior to the observed TeV flare, a flare in both HE  \citep[\textit{Fermi}-LAT,][]{Abdo2011} and X-ray \citep[MAXI,][]{Isobe2010} was observed (without simultaneous VHE observations). This HE/X-ray flare triggered the VHE observations.
%%%%%%%%%%%%%%%%%%%%%%%%%%%%%%%%%%%%%%%%%%%%%%%%%%%%%%%
\paragraph{MAXI.} MAXI is an all-sky monitoring instrument onboard the International Space Station, and is sensitive to X-rays in the energy range 0.5--30 keV \citep{Matsuoka09}. 
We downloaded the daily-binned light curve for the entire month of February from the MAXI Science Center data archive\footnote{http://maxi.riken.jp}.  Mrk\,421 is bright enough to result in a significant detection in each 24-hour time bin.  The resulting light curve, presenting the 2--20 keV count rate in daily time bins, is shown in the third panel from the top in Figure \ref{fig:mwlLC}. 

%%%%%%%%%%%%%%%%%%%%%%%%%%%%%%%%%%%%%%%%%%%%%%%%%%%%%%%
\paragraph{\textit{RXTE}-PCA.}
We observed Mrk\,421 with the Proportional Counting Array (PCA) instrument onboard \textit{RXTE} through two observing programs \added{(Obs IDs: 95386, 95133)}.  
A total of 42 \textit{RXTE} observations were carried out between 1 February, 2010 and 1 March, 2010. The \textit{RXTE} datasets relevant for this paper are shown in the fourth panel from the top in Figure \ref{fig:mwlLC} binned by individual observations.

For each observation, we extracted the spectrum from the Standard-2, binned-mode data (i.e.\,129 channel spectra accumulated every 16 seconds) using \texttt{HEASoft} v6.11.  We screened the data so that the angular separation between Mrk\,421 and the pointing direction was less than 0.05$^{\circ}$, the elevation angle was greater than 5$^{\circ}$, the time since the last passage of the South Atlantic Anomaly was greater than 25 minutes, and the electron contamination was low ({\tt ELECTRON2} $<$ 0.1).  We estimated the background using the L7 model for Epoch 5C---the Proportional Counting Unit (PCU) count rate of Mrk\,421 was close to the transition point where the bright background model is recommended over the faint background (40 cts/s/PCU). We chose the background model based on the observed mean count rate in each observation.  All spectra were accumulated from both anodes in the upper Xenon layer of PCU2, which is turned on in every observation and was the only PCU in operation in most of our observations.  Since the PCA has low sensitivity below 2.5 keV, and Mrk\,421 is faint above 20 keV, we analyzed the background-subtracted spectra in the energy range 2.5-20 keV.

To enable a more careful study of the X-ray behavior as well as joint X-ray/VHE $\gamma$-ray behavior during the decline phases described in Section~\ref{VHE-xray-Correlation}, we produced more-detailed light curves for observations taken during the P95133 period (see 
Figures~\ref{fig:xray-VHE-ilc} and \ref{fig:ii-ff-xray-VHE}).  We determined the \textit{RXTE} count rate with the REX analysis tool\footnote{\url{http://heasarc.gsfc.nasa.gov/docs/xte/recipes/rex.html})} using the same extraction criterion as above. Light curves were first extracted in 16 second time bins and then re-binned using the \emph{ftool} {\tt lcurve} to create 10-minute time bins.

%%%%%%%%%%%%%%%%%%%%%%%%%%%%%%%%%%%%%%%%%%%%%%%%%%%%%%%
\paragraph{\textit{Swift}-XRT.}
We analyzed \textit{Swift} observations of Mrk\,421 from two observing programs: 31630 and 30352 (the latter initiated in response to the VHE flare).  A total of 23 observations were carried out between 1 February, 2010 and 1 March, 2010. The light curve from these observations is shown in the fourth panel from the top in Figure \ref{fig:mwlLC} binned by individual observations.  Due to the high count rate of the source ($>$20 cts/s) all observations were obtained in windowed timing (WT) mode. We reran the \textit{Swift} data reduction pipeline on all datasets ({\texttt xrtpipeline} v0.12.6) to produce cleaned event files and exposure maps.  We created source spectra using \texttt{XSelect} v2.4b, extracting source events from a circular region of radius 40$''$ centered on the source.  We subsequently created ancillary response files using \texttt{xrtmkarf} v0.5.9, applying a PSF and dead-pixel correction using the exposure map created with \texttt{xrtpipeline}.  Finally, the appropriate response matrix file (in this case swxwt0to2s6\_20010101v013.rmf) was taken from the \textit{Swift} calibration database.  We grouped the spectra to have a minimum of 20 counts per bin in order to facilitate the use of $\chi^2$-statistics in \texttt{Xspec} and carried out model fits in the 0.3--10 keV energy range.

%%%%%%%%%%%%%%%%%%%%%%%%%%%%%%%%%%%%%%%%%%%%%%%%%%%%%%%%%%%%%%%%%%%%%%
\subsection{Optical Observations}
\label{DataOptical}

\paragraph{UVOT.}
The Ultraviolet/Optical Telescope onboard \textit{Swift} also obtained data
during each observation, in one of three UV filters (UVW2, UVM2 or
UVW1) for a total of 59 exposures. All of the data taken between 7 February, 2010 and
20 February, 2010 were analyzed and are shown in the fifth panel from the top of Figure~\ref{fig:mwlLC} binned by individual observations.  
After extracting the source counts from an aperture of 5.0$''$ radius
around Mrk\,421 and the background counts from four neighboring
regions, each of the same size, the magnitudes were computed using the
\texttt{uvotsource}\footnote{HEASOFT v6.13,
  Swift$\_$Rel4.0(Bld29)$\_$14Dec2012 with calibrations from
  \citet{2011AIPC.1358..373B}.} tool. These were converted to fluxes
using the central wavelength values for each filter from \citet{2008MNRAS.383..627P}. The observed fluxes were corrected for Galactic extinction following the procedure and R$_v$ value in \citet{2009ApJ...690..163R}. An $E(B-V)$ value of 0.013
from \citet{2011ApJ...737..103S} was used.

\paragraph{Ground-based Optical Observatories.}
The optical fluxes reported in this paper were obtained within the
GASP-WEBT program \citep[e.g.][]{Villata2008, Villata2009}, with the optical telescopes at
Abastumani, Roque de los Muchachos (KVA), Crimean, and Lowell (Perkins) observatories.
Additional observations were performed with the Goddard Robotic Telescope (GRT), Galaxy View, and New Mexico Skies. 
All instruments used the calibration stars reported
in \citet{Villata1998} for calibration. The Galactic extinction was
corrected with the reddening corrections given in \citet{schlegel1998}. The flux from the host galaxy (which is significant only below $\nu \sim 10^{15}$~Hz) was estimated using the flux values across the R band from \cite{Nilsson2007} and the colors reported in \cite{Fukugita1995}, and subsequently subtracted from the measured flux. 
The flux values obtained by the various observatories during the same 6 hour time period agree within uncertainties except for the GRT, which shows a flux systematically 15\% lower than that of the other telescopes. We therefore assume this represents a systematic error in the data, and correct the observed fluxes to match the fluxes from the other observatories  during the same 6-hour time interval.

Additionally, \added{high-cadence, 2-minute exposure} optical R-band observations near-simultaneous with VERITAS were obtained on 17 February, 2010 with the 1.3-m Robotically Controlled Telescope (RCT) located at Kitt Peak National Observatory.  
The RCT observations started $\sim$50 minutes after the beginning of the VERITAS observations and ended $\sim$15 minutes after VERITAS stopped observing.

The reported fluxes from all optical observatories include instrument-specific offsets of a few mJy.  These are due to differences in filter spectral responses and analysis procedures of the various optical data sets combined with the host-galaxy contribution (about 1/3 of the total flux measured for Mrk\,421 in the R band). The following offsets were determined and corrected for by using simultaneous observations and treating several of the GASP-WEBT instruments as reference: GRT = 2.5~mJy; RCT = -1.0~mJy; CRAO = 3.0~mJy;  RCT for the long observations on 17 February = 4.0~mJy. Moreover, a point-wise fluctuation of 0.2 mJy ($\sim$0.01 mag) was added in quadrature to the statistical uncertainties in order to account for potential day-to-day differences for observations with the same instrument. 

The reconstructed optical fluxes 
are shown in the sixth panel from the top of Figure \ref{fig:mwlLC} binned by individual observations. The 2-minute-binned VERITAS and RCT light curves are displayed in Figure~\ref{fig:veropt_lcs}; these latter light curves are used in the discrete cross-correlation analysis detailed in Section~\ref{OpticalCorrelation}.
%%%%%%%%%%%%%%%%%%%%%%%%%%%%%%%%%%%%%%%%%%%%%%%%%%%%%%%

%%%%%%%%%%%%%%%%%%%%%%%%%%%%%%%%%%%%%%%%%%%%%%%%%%%%%%%
\paragraph{Steward Observatory - Optical Polarization.}

Optical observations of Mrk\,421 were made during the high-energy monitoring campaign by the Steward Observatory (SO) 2.3m Bok Telescope on Kitt Peak, Arizona. The source was observed on seven consecutive nights from 13 February, 2010 (MJD 55240) through 19 February, 2010 (MJD 55246) using the SPOL imaging/spectropolarimeter~\citep{Schmidt1992b}. On all seven nights, flux and polarization spectra spanning 4000-7550 \AA\ were acquired  using a 600 lines/mm grating in first order, which gives a dispersion of 4~\AA\,pixel$^{-1}$. The polarization measurements employed a $3$ arcsec-wide slit, yielding a resolution of $\sim$20~\AA.  The slit length was $51''$, long enough to sample the sky background from a region without a significant amount of light from the host elliptical galaxy of Mrk\,421~\citep{Ulrich75}.  Data reduction followed the same general procedure as outlined in \citet{Smith2003}.  The bottom panel of Figure~\ref{fig:mwlLC} shows the results of the spectropolarimetry averaged over a 2000~\AA-wide bin centered at 6000~\AA. The broad-band polarization measurements were not corrected for the unpolarized starlight from the host galaxy of Mrk\,421 falling within the $3''\times10''$ spectral extraction aperture. Such a correction would increase the level of optical polarization, but does not affect the measured polarization position angle. In addition to the spectropolarimetry, differential spectrophotometry was acquired with a slit width of 7.6$''$.  Again, a 10$''$-wide extraction aperture was used for both Mrk\,421 and a comparison star calibrated by~\citet{Villata1998}. No correction for the host galaxy was made to the reported R magnitudes since the AGN still dominates the total flux observed in the larger aperture. The largest flux variation observed in Mrk 421 during this period was $\sim$0.1 mag between 17 February (MJD 55244) and 19 February, 2010 (MJD 55246). 

%%%%%%%%%%%%%%%%%%%%%%%%%%%%%%%%%%%%%%%%%%%%%%%%%%%%%%%%%%%%%%%%%%%%%%
\subsection{Radio Observations \label{DataRadio}}

Contemporaneous radio data were taken with 
the 40-m Owens Valley Radio Observatory (OVRO) telescope at 15 GHz, 
the 26-m University of Michigan Radio Astronomy Observatory (UMRAO) at 14 GHz and 8 GHz, and the 
14-m Mets\"ahovi Radio Observatory at 37 GHz. Details of the observing
strategy and data reduction are given by \citet[OVRO]{Richards2011};
\citet[UMRAO]{Aller1985}; and  \citet[Mets\"ahovi]{Terasranta1998}.
Mrk\,421 is a point-like source for the three above-mentioned single-dish radio instruments,
which means that the measured fluxes are the flux densities integrated
over the full source extension. The light curves are shown in the second from bottom panel of Figure \ref{fig:mwlLC} binned by individual observations.

%%%%%%%%%%%%%%%%%%%%%%%%%%%%%%%%%%%%%%%%%%%%%%%%%%%%%%%%%%%%%%%%%%%%%%
%%%%%%%%%%%%%%%%%%%%%%%%%%%%%%%%%%%%%%%%%%%%%%%%%%%%%%%%%%%%%%%%%%%%%%
%%%%%%%%%%%%%%%%%%%%%%%%%%%%%%%%%%%%%%%%%%%%%%%%%%%%%%%%%%%%%%%%%%%%%%
\section{Results from  the Exceptional Flare on MJD 55244} \label{tevresults}
%%%%%%%%%%%%%%%%%%%%%%%%%%%%%%%%%%%%%%%%%%%%%%%%%%%%%%%%%%%%%%%%%%%%%%
The flux state observed by VERITAS during the 17 February, 2010 (MJD 55244) flare is extraordinary; it is the highest flux state in Mrk\,421 ever observed in VHE $\gamma$-rays. The peak fluxes measured above specific energy thresholds are given as  $\sim 11$ CU above 110~GeV, $\sim 15$ CU above 200~GeV , $\sim 17$ CU above 420~GeV, $\sim 21$ CU above 600~GeV, $\sim 27$ CU above 1~TeV; the higher flux with higher threshold energy is due to Mrk\,421 exhibiting a much harder spectrum during the flare than the Crab Nebula. 
The ``baseline" average flux from the MAGIC data just prior to the main flare is $2.4\times 10^{-10}\, \mbox{photons}\, \mbox{cm}^{-2}\, \mbox{s}^{-1}$ above 200 GeV, which is just below 1 CU.  
The VERITAS $\gamma$-ray and RCT R-band optical data are the only two bands to have high sampling rates during the night of this exceptional VHE flare.  In this section, we detail the results from the VERITAS observations along with the optical-VHE correlation analysis.

%%%%%%%%%%%%%%%%%%%%%%%%%%%%%%%%%%%%%%%%%%%%%%%%%%%%%%%%%%%%%%%%%%%%%%
\subsection{Temporal Variability in the VHE $\gamma$-Ray Band} \label{VHESpectra}
%%%%%%%%%%%%%%%%%%%%%%%%%%%%%%%%%%%%%%

\begin{figure}[ht]
\epsscale{1.15}
\plotone{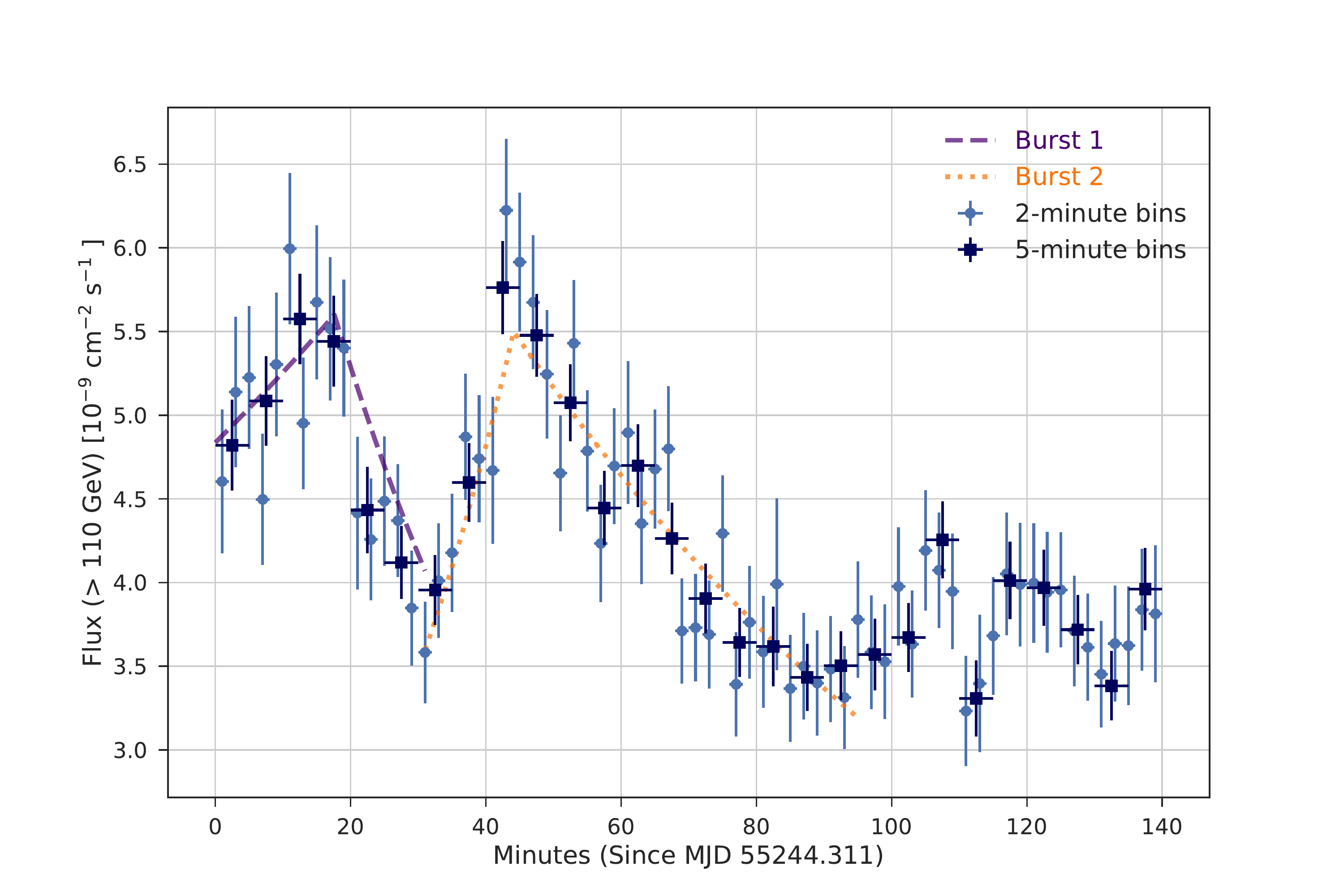}
\caption{Light curve (2-minute and 5-minute bins) for VERITAS Mrk\,421 data above $110$ GeV for the first 2.33 hours of observations on MJD 55244 where two ``bursts" are identified via a Bayesian Block analysis.} The dashed lines show an \emph{Exponential (Exp)} function  fit to the rise and fall of the two bursts using the 2-minute-binned light curve. The fit parameters, including the rise and decay times are provided in Table~\ref{tab:burstfits}.
\label{fig:ver_expfits}
\end{figure}

%%%%%%%%%%%%%%%%%%%%%%%%%%%%%%%%%%%%%%%%%%%%%%%%%%%%%%%
The high-statistics VERITAS data of the giant Mrk\,421 flare enables construction of finely-binned light curves. The energy threshold depends on the elevation of the observations, increasing for smaller source elevation angles. 420~GeV represents the lowest  energy threshold common to the $\sim$5 hours of data taken during the night of the flare (the full-night 2-minute-binned light curve with this threshold is shown in Figure~\ref{fig:veropt_lcs}). For the first $\sim$2.33 hours of the night, the elevation of the source was above $ 75^{\circ}$, resulting in an energy threshold of 110~GeV for light curves generated with these data. For this part of the flare night ($\sim$140 minutes), we constructed 2-minute and 5-minute-binned light curves above $110~\text{GeV}$ (shown in Figure~\ref{fig:ver_expfits}) to characterize any strong variability as discussed directly below. In addition, we constructed 2-minute-binned light curves for three energy bands with equal statistics in each band: a ``low-energy'' band, defined as $110~\text{GeV} < E < 255~\text{GeV}$;  a ``medium-energy'' band, defined as $255~\text{GeV} < E < 600~\text{GeV}$; and a $\geq 600$~GeV ``high-energy'' band. We investigate the fractional variability for these three bands in Section \S\ref{fvar}. 

%%%% FIGURE 2 %%%%%%%%

%%%% TABLE 1 %%%%%%%%

\begin{deluxetable}{c|l|cccccc}[!htbp]
\tablecaption{Results from fits to the 2-minute light curves for the two bursts shown in Figure~\ref{fig:ver_expfits}. The quoted (most likely) values represent the 50$^{th}$ percentile, while the uncertainties are given as the 90$^{th}$ percentile of the posterior distributions of the parameters. $t_{\text{rise}}$ and $t_{\text{decay}}$ are the doubling and halving timescales, respectively. The units for the normalization $A$ are [10$^{-9}$ photons cm$^{-2}$ s$^{-1}$ TeV$^{-1}$]; $\kappa$ is unitless.
\label{tab:burstfits}}
\tablehead{Burst & Fit & \colhead{$A$} & \colhead{t$_{\text{rise}}$} & \colhead{t$_{\text{decay}}$} & \colhead{t$_{peak}$} & \colhead{$\kappa$}  & \colhead{$\chi^{2}$/NDF} \\
 & Function & & \colhead{[min]} & \colhead{[min]} & \colhead{[min]} & }
\startdata
1 & Exp & 5.5$^{+0.34}_{-0.28}$ & 84$^{+\text{$\infty$}}_{-49}$ & 28$^{+20}_{-9.4}$ & 18$^{+3.4}_{-5.2}$ &  -- &18/12 \\
& GG & 5.8$^{+2.3}_{-0.6}$ & 180$^{+63}_{-100}$ & 55$^{+91}_{-23}$ &  17$^{+2.6}_{-5.2}$ & 0.64$^{+0.31}_{-0.41}$  &11/11\\
\hline
\hline
2 & Exp & 5.5$^{+0.26}_{-0.26}$ & 22$^{+13}_{-6.7}$ & 65$^{+13}_{-9.6}$ & 44$^{+2.3}_{-2.1}$& -- &30/29\\
& GG &6.6$^{+1.6}_{-0.88}$ & 30$^{+30}_{-18}$ &78$^{+40}_{-26}$& 44$^{+2.0}_{-1.3}$ & 0.47$^{+0.28}_{-0.19}$ &  27/28 \\
\enddata
\end{deluxetable}

%%%% FIGURE 3 %%%%%%%%

\begin{figure}[ht]
\epsscale{1.15}
\plottwo{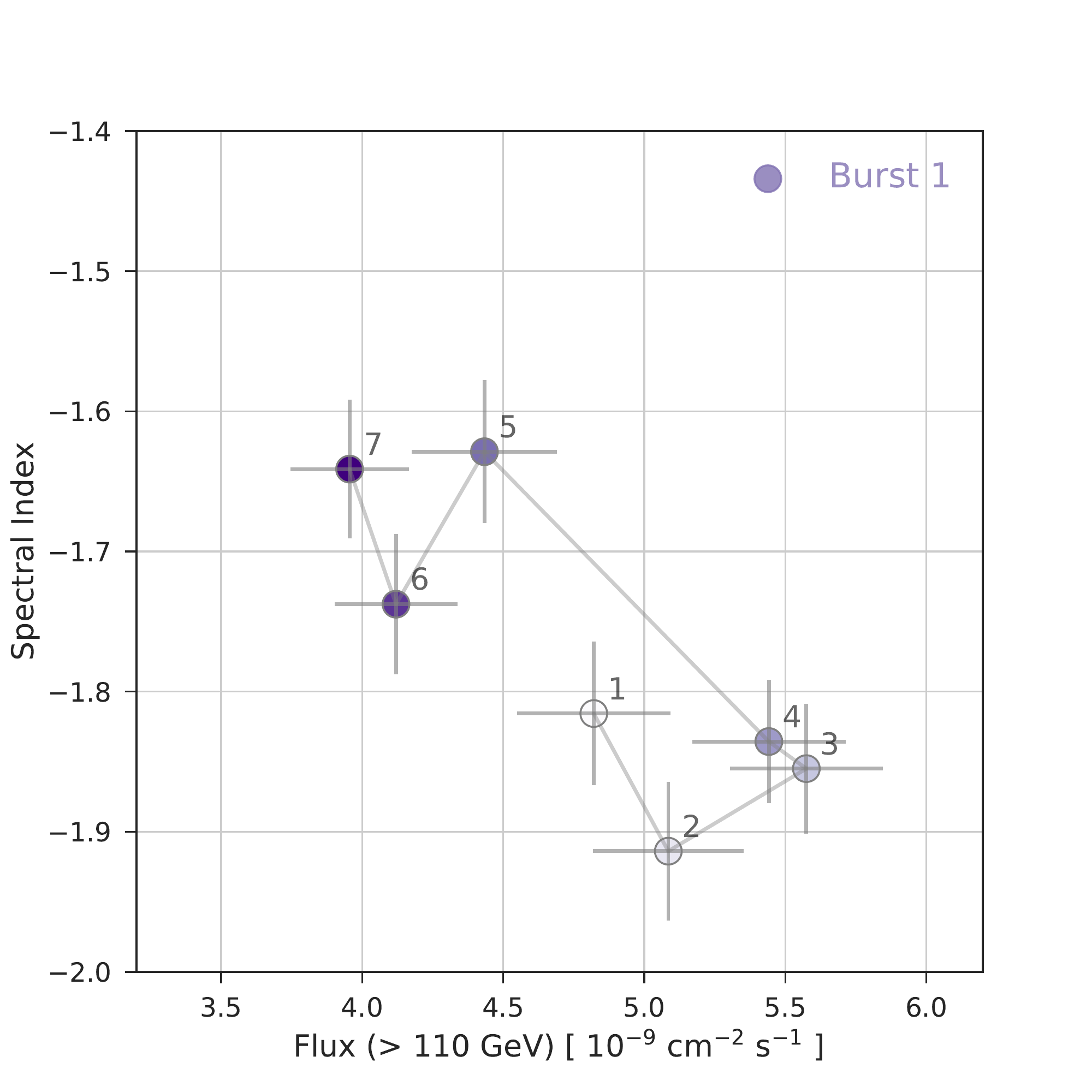}{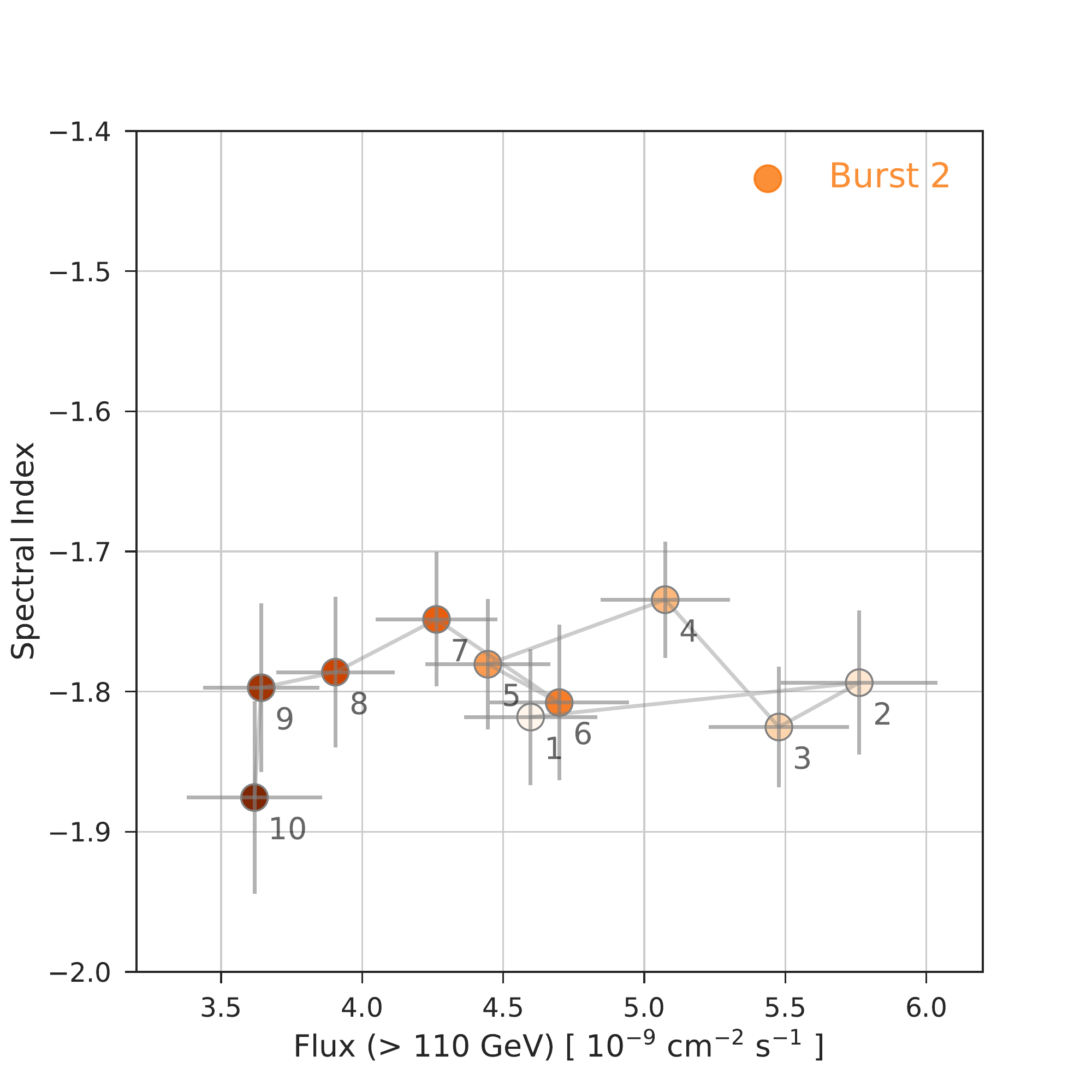}
\caption{Photon index versus flux of the VERITAS detections of Mrk\,421 over 5-minute intervals shown separately for \emph{Burst 1} (\textit{left}) and \emph{Burst 2} (\textit{right}). The color shades represent the chronological progression of the bursts, with lighter colors corresponding to earlier times. The indices are obtained by a fit with an exponential cut-off power law ($E_{cut}$), where $E_{cut}$ is fixed to the global value of 4 TeV.
\label{fig:ver_hysts}}
\end{figure}

Figure~\ref{fig:mwlLC} shows the full set of VHE data during February, 2010 binned in nightly bins.   
The first observations by VERITAS on 17 February, 2010  
are likely to have been taken after the onset of the flare;
thus, we cannot make any statement about the rise-time of the main flare. A decay function ($N(t)=N_0\cdot 2^{-t/t_{\text{decay}}}$, where $t_{\text{decay}}$ is the halving time) was fit to the four VERITAS data points in Figure~\ref{fig:mwlLC} resulting in a halving timescale of $\sim$1 day. We fit the same function to the 10-minute binned data resulting in a halving time of $1.17 \pm 0.07$ days, consistent with the nightly-binned result.

A Bayesian Block analysis \citep{Scargle2013} was applied to the $>110$~GeV VERITAS data from the flare night to look for any significant change points. Two peaks or ``micro-bursts'' were identified in this manner in the first $\sim$140 minutes. Figure~\ref{fig:ver_expfits} shows a zoom in on this region with the peaks clearly visible in the first $\sim$95 minutes. \emph{Burst 2} shows an apparent asymmetry which can be quantified via the method used in \cite{Abdo2010}; the symmetry parameter is found to be $\xi = 0.50 \pm 0.09$ corresponding to moderate asymmetry. We do not quote the asymmetry value for \emph{Burst 1} as we cannot be certain we observed the full rise of the burst.  
In addition, we fit several functions to these data to determine the most likely rise and decay timescales for the peaks. We test an \emph{Exponential (Exp)}, F$(t) = A\cdot exp{\left[|t-t_{\text{peak}}|/t_{\text{rise,decay}}\right]}$, and the \textit{Generalized Gaussian (GG)} burst profile from~\citet{Norris1996} of the form, F$(t) = A\cdot exp{\left[|t-t_{\text{peak}}|/t_{\text{rise,decay}}\right]^{\kappa}}$. The most likely values and uncertainty of the parameters are determined using a Markov Chain Monte Carlo (MCMC) method with the \texttt{emcee} tool~\citep{emcee}. The most likely values are taken as the 50$^{th}$ percentiles, while the uncertainties are given as 90\% confidence intervals of the posterior distributions of the parameters. The fit results are provided in Table~\ref{tab:burstfits}. For \emph{Burst 2}, the \emph{Generalized Gaussian} profile with one more parameter than the \emph{Exponential} function is not statistically preferred. In Section~\ref{conclusions} we use the \emph{Burst 2} rise time, $t_{rise}=22$~minutes to place an upper bound on the effective size of the emission region as well as a lower bound on the Doppler factor when taking into account compactness and opacity requirements of the emitting region.

%%%%%%%%%%%%%%%%%%%%%%%%%%%%%%%%%%%%%%%%%%%%%%%%%%%%%%%
%VHE hysteresis study during flare
%%%%%%%%%%%%%%%%%%%%%%%%%%%%%%%%%%%%%%%%%%%%%%%%%%%%%%%%%%%%%%%%%%%%%%
\subsection{Search for VHE hysteresis during flare}
In order to investigate possible relationships between flux and photon index for the $>110$~GeV VERITAS data, coarser 5-minute bins were used for reducing statistical uncertainties. Within each time bin, a flux estimation and a full spectral reconstruction were performed. The Mrk\,421 spectra are curved within each 5-minute bin, therefore an exponential cutoff power-law function: 

\begin{equation}
\frac{dN}{dE} = I_{0}\left(\frac{E}{1~\text{TeV}}\right)^{\Gamma} exp{\left(-\frac{E}{E_{cut}}\right)},
\end{equation}
was used to reconstruct and fit the spectra, where $E_{cut}$ is the cutoff energy. The ${E_{cut}}$ parameter was kept fixed to ${E_{cut}}=4$~TeV, the value from a global fit. Figure~\ref{fig:ver_hysts} displays the resulting photon index versus flux representation of the VERITAS detections of Mrk\,421 for the two identified bursts.
While there is some evidence for a counter-clockwise hysteresis loop or a softer-when-brighter trend for \textit{Burst 1}, the photon index in \textit{Burst 2} is very stable while the flux rises and falls by a factor of $\sim$1.5. 

The index-flux relationship for \textit{Burst 1} was assessed with simple $\chi^{2}$ tests using the observed quantities against constant and linear models as the null hypotheses. A constant model can be rejected with a p-value of $3.2\times10^{-5}$ ($\chi^{2}$/NDF = 30/6), while the p-value for a linear model is 0.07 ($\chi^{2}$/NDF = 10/5). A $\chi^{2}$ difference test prefers the linear model over the constant model with a p-value of $6.8\times10^{-6}$. 
To test the statistical robustness of this relationship, the observed data points were resampled within the measurement uncertainties and the $\chi^{2}$ tests were repeated for 100,000 iterations. At a 90\% confidence level, the constant model is rejected 99.7\% of the time, while the linear model is rejected 81.9\% of the time. The p-value for the linear model indicates the index-flux relationship for \textit{Burst 1} appears only marginally consistent with a linear (softer-when-brighter) trend. The deviation from a linear trend could be an indication of a more complicated relationship between the Mrk 421 index and flux, such as a hysteresis loop.

These index-flux characteristics along with the asymmetry of \textit{Burst 2} can be used to infer differing relationships between the cooling and acceleration timescales for the bursts, which is further discussed in Section~\ref{conclusions}.

%%%%%%%%%%%%%%%%%%%%%%%%%%%%%%%%%%%%%%%%%%%%%%%%%%%%%%%
%Optical-VHE correlation studies
%%%%%%%%%%%%%%%%%%%%%%%%%%%%%%%%%%%%%%%%%%%%%%%%%%%%%%%%%%%%%%%%%%%%%%
\subsection{VHE $\gamma$-Ray and Optical Correlation Studies.} \label{OpticalCorrelation}

%%%%%%%%%%%%%%%%%%%%%%%%%%%%%%%%%%%%%%%%%%%%%%%%%%%%%%%

\begin{figure}[ht]
\epsscale{1.15}
\plotone{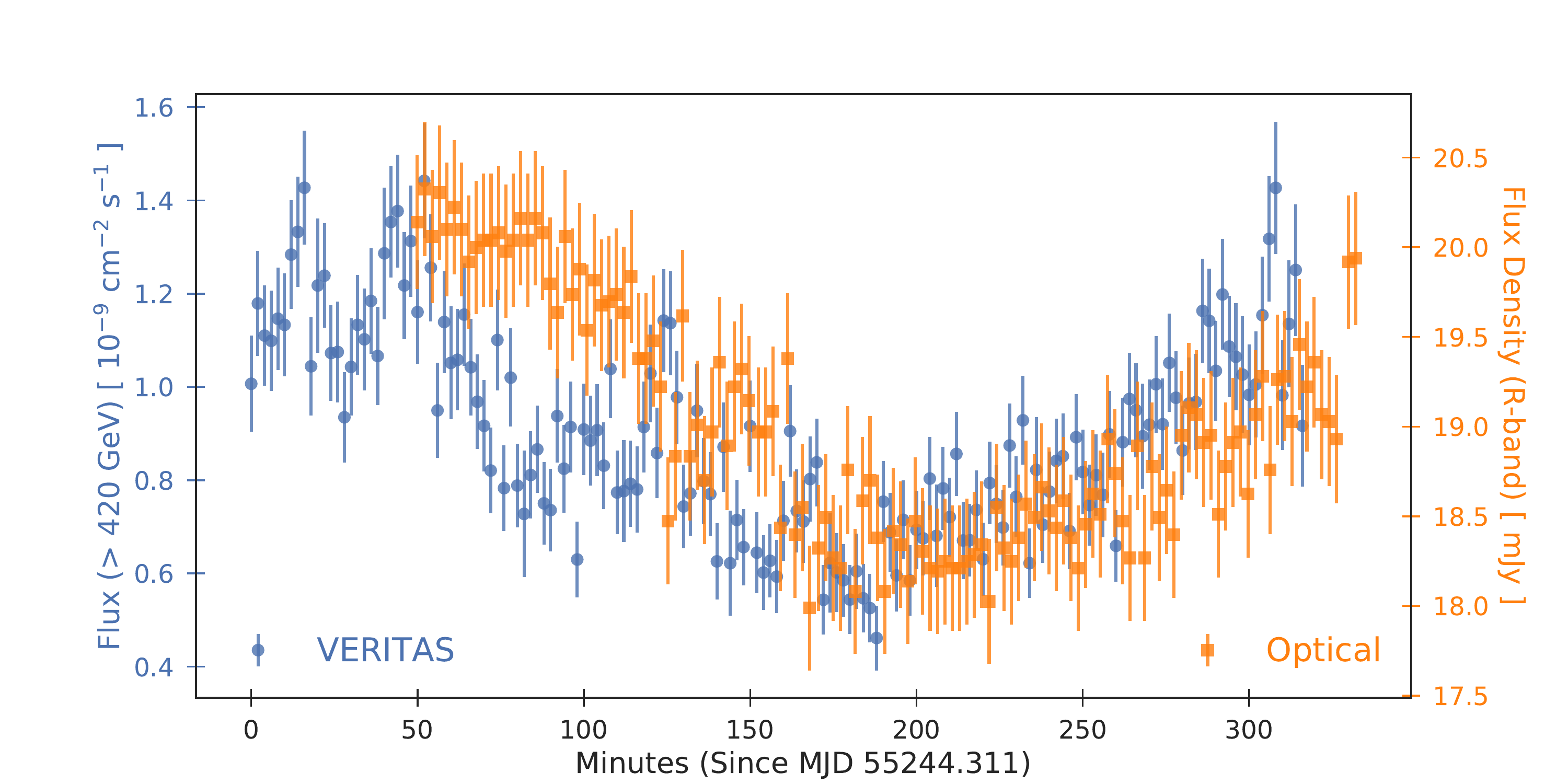}
\caption{2-minute-binned VERITAS $> 420$~GeV (blue) and RCT optical R-band (orange) light curves during  MJD 55244. \label{fig:veropt_lcs}}
\end{figure}

The RCT R-band optical data are the only dataset other than VERITAS to have high statistical sampling during the night of the VHE highest state (MJD 55244). 
Figure~\ref{fig:veropt_lcs} shows the VERITAS 2-minute-binned data (blue) over the full energy range ($420~\text{GeV} <E<30~\text{TeV}$) commensurate with the low-elevation threshold, and the $R$-band optical data (orange) overlaid. Visual inspection indicates an apparent correlation between the two wavebands which warrants further investigation.

%%%% FIGURE 4 %%%%%%%%

\begin{figure}[h]
\epsscale{1.15}
\plotone{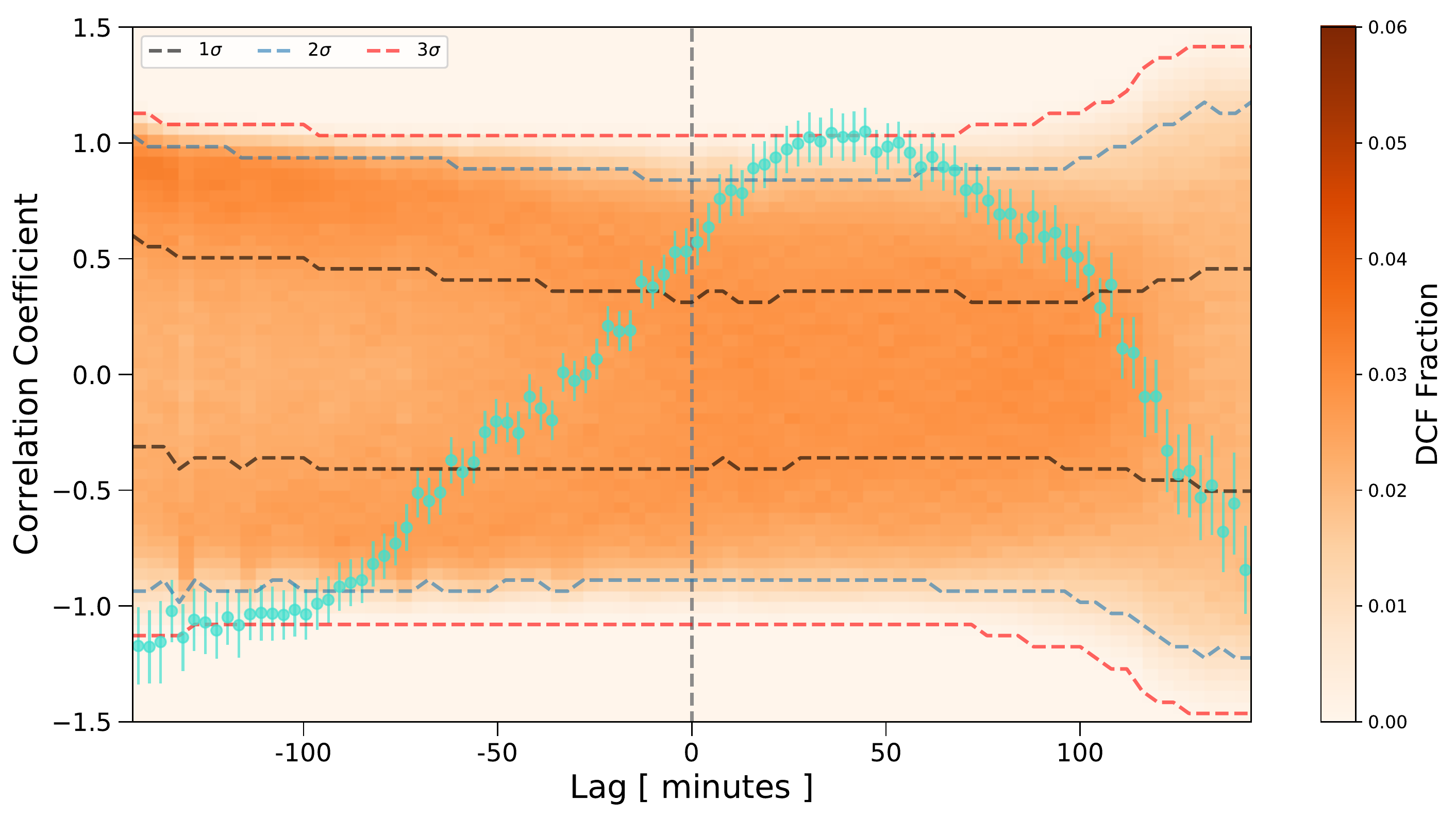}
\caption{Simulated DCCFs for VERITAS ($>420$~GeV) and optical R-band light curves. The DCCF from observations is in turquoise. ``DCF fraction'' represents the fraction of times a simulated DCCF falls in a given lag-time and correlation-coefficient bin (shown with the 2D histogram color map). The DCF fraction histogram (representing a PDF) is integrated to obtain the confidence levels. The black, blue, and red dashed lines show the 1$\sigma$, 2$\sigma$, and 3$\sigma$ levels respectively. A positive lag time corresponds to a delay in the optical light curve with respect to the VERITAS light curve.\label{fig:mrk421_dccf}}
\end{figure}

We used the discrete cross-correlation function (DCCF) analysis following~\citet{edelson90} to test for time lags between the 2-minute-binned VERITAS and RCT light curves. The DCCF was calculated after subtracting the mean from each light curve and dividing the result by the standard deviation. There is a broad peak apparent in the DCCF (turquoise points in Figure~\ref{fig:mrk421_dccf}) centered at a lag time of roughly 45~minutes, with VHE $\gamma$-rays leading the optical. 

In order to assess the statistical significance of features in the DCCF, including the broad peak, Monte Carlo simulations were performed following the method by~\citet{emmanoulopoulos13}.
First, as described in Appendix~\ref{sec:psds}, the Power Spectral Density (PSD) was constructed and fitted for both the VERITAS and optical light curves. Next, the best-fit VERITAS PSD  ($P(f) \propto f^{-1.75}$) was used to generate 100,000 random light curves. The random light curves were then paired with the \textit{observed} optical light curve to calculate a DCCF for each iteration. 

%%%% FIGURE 5 %%%%%%%%

 Figure~\ref{fig:mrk421_dccf} shows the resulting simulated DCCFs binned into a 2D histogram of correlation coefficient versus lag time  bins. The bin contents of the 2D histogram are normalized such that for a fixed lag time, each correlation coefficient bin gives the fraction of all DCCFs falling within the bin, and the bin contents along the vertical axis will sum to one. Significance levels are estimated by integrating the probability density function (PDF) represented by the 2D histogram of simulated DCCFs. The VERITAS-optical DCCF shows evidence for a signal at lag times of 25--55 minutes. The significance of the correlation is $\sim3\sigma$. 
The use of an observed light curve (in this case, the optical R-band) in the significance level estimation is a conservative approach. If simulated light curves are generated from the optical PSD ($P(f) \propto f^{-1.85}$), the correlation significance increases to $\sim 4 \sigma$. We note, however, that the PSD fit errors are large, hindering a good characterization of the uncertainties on the significance of the correlation.

%%%%%%%%%%%%%%%%%%%%%%%%%%%%%%%%%%%%%%%%%%%%%%%%%%%%%%%%%%%%%%%%%%%%%%
%%%%%%%%%%%%%%%%%%%%%%%%%%%%%%%%%%%%%%%%%%%%%%%%%%%%%%%%%%%%%%%%%%%%%%
%%%%%%%%%%%%%%%%%%%%%%%%%%%%%%%%%%%%%%%%%%%%%%%%%%%%%%%%%%%%%%%%%%%%%%

%%%%%%%%%%%%%%%%%%%%%%%%%%%%%%%%%%%%%%
\subsection{Autocorrelation Analysis with the VHE Flare} \label{AutoTeV}
%%%%%%%%%%%%%%%%%%%%%%%%%%%%%%%%%%%%%%%%%%%%%%%%%%%%%%%%%%%%%%%%%%%%%%
The VHE flux from Mrk\,421 shows clear intra-night variability during the night of the flare on 17 February, 2010 and the PSD analysis for VERITAS from Appendix~\ref{sec:psds}  
shows a power spectrum of pink noise (or flicker noise) with $P(f) \propto f^{-1.75}$.  
However, these results are limited to shortest timescales of $\sim500$ seconds.

A modified auto-correlation function (MACF) proposed in \citet{Li2001} and extended in \citet{Li2004} could provide improved sensitivity to short variations of the VHE flux. 
Details of the method can be found in Appendix~\ref{appendix_MACF}. 
Though \citet{Bolmont2009} used the method to search for signatures of potential Lorentz invariance violation (LIV)  in the 2006 PKS 2155-304 flare, it is a  novel technique for VHE variability studies.
In Appendix~\ref{appendix_MACF}, we have applied the MACF to the night of the VHE flare (Epoch 3; see Section~\ref{LightCurves}) using all events above an energy threshold $E=420$~GeV.
No critical timescale is observed on these short timescales, but the data are consistent with a stochastic process or ``pink noise"  corroborating the VERITAS  PSD results found at longer timescales in Appendix~\ref{sec:psds}. Probed by the combination of these two techniques, this is the first time that this stochastic behavior has been shown to exist in a blazar on the full range of timescales from  seconds to hours.

%%%% FIGURE 6 %%%%%%%%

\begin{figure}[!htbp]
\centering{
\includegraphics[width=1.0\textwidth]{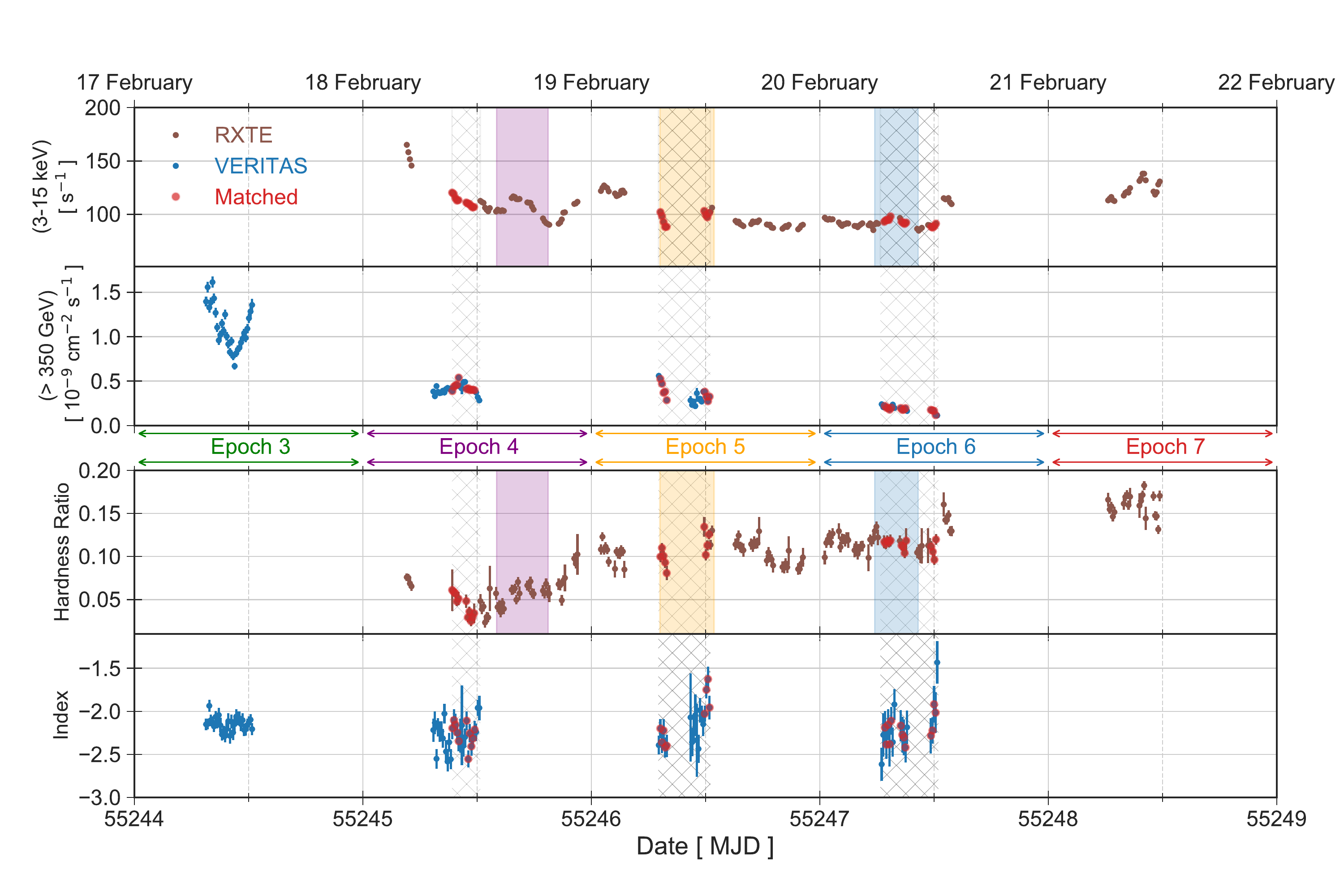}
}
\caption{\label{fig:xray-VHE-ilc} Detailed 10-minute-binned light curves for VHE (VERITAS - blue) and X-ray (\textit{RXTE} - brown) data for Epochs 3-7: the VHE flare (Epoch 3) followed by the three VHE decline epochs (Epochs 4-6) and a final epoch during which \textit{RXTE} count rates are elevated again.
\textit{Matched} - red points distinguish data where there is strictly simultaneous overlap between \textit{RXTE} and VERITAS observations. 
The top panel shows as a function of time, the \textit{RXTE} count rate and VERITAS flux light curves, while the bottom panel shows the \textit{RXTE} hardness ratio between the 5-15~keV and 3-5~keV bands and the VERITAS photon index from power-law fits between 350~GeV and 3~TeV.
We note that there are no simultaneous X-ray data during
the VHE flare (Epoch 3) and there are no VHE data during Epoch 7. Regions of overlap are indicated by gray hatches and their behavior studied in Section ~\ref{VHE-xray-Correlation}. Color-shaded regions are used for a more in-depth X-ray hardness ratio-count rate study illustrated by the bottom panels of Figure~\ref{fig:rxte_hysts_epochs}.}
\end{figure}

\section{Results from Full February 2010 Multiwavelength Data Set} \label{mwlresults}

%%%%%%%%%%%%%%%%%%%%%%%%%%%%%%%%%%%%%%%%%%%%%%%%%%%%%%%%%%%%%%%%%%%%%%
\subsection{Light Curves} \label{LightCurves}
In this section, we focus on the multiwavelength light curves of Mrk\,421 
for February, 2010. Figure~\ref{fig:mwlLC} shows the light curves for each waveband participating in the campaign. The VHE data are shown averaged over the full set of observations for a given night spanning durations between $\sim$twenty minutes and $\sim$six hours;  \emph{Fermi}-LAT and MAXI data are shown with daily binning; all other light curves are binned by individual exposures.
To study the flux properties of the VHE data in more detail, the entire combined MAGIC and VERITAS dataset from Figure \ref{fig:mwlLC} was split into multiple epochs.   
MAGIC data are available for several days leading up to the flare. These Epoch 1 data (MJD 55232-55240)  are used as ``baseline" VHE data to which we compare the flaring period and its decline. Epoch 2 (MJD 55240-55243) has no VHE data, however it is used to study the behavior of the X-ray and HE data as the flare builds up in these bands (see Section \ref{HE-xray-Correlation}). 
Epoch 3 comprises the main flare (MJD 55244) showing extraordinary overlap between the VHE and optical data enabling the correlation analysis shown in Section~\ref{OpticalCorrelation}. 
Epochs 3-7 (MJD 55244, 55245, 55246, 55247, 55248) are shown in Figure~\ref{fig:xray-VHE-ilc} which displays 10-minute-binned light curves for both VHE and \emph{RXTE} X-ray data (top two panels).  Epochs 3-6  comprise only  VERITAS data in the VHE band (shown above 350 GeV as the lowest common threshold) and are, respectively, during the VHE flare and just afterwards in three decline epochs. Epochs 4-7 comprise \emph{RXTE} data in the 3-15~keV band where Epochs 3-6 overlap with the VERITAS data during the decline phases and a subsequent rise in \emph{RXTE} data  is seen in Epoch 7; no VHE data are available in this last epoch. During periods where strictly simultaneous data were obtained, we matched the start and stop times of each time bin between the VERITAS and \textit{RXTE} light curves. These VHE and X-ray light curves, along with the VERITAS photon indices and \emph{RXTE} hardness ratios shown in the bottom two panels of Figure~\ref{fig:xray-VHE-ilc} are used in more detailed studies in Section \ref{VHE-xray-Correlation}. However, first we compare variability properties across all participating wavebands shown in Figure~\ref{fig:mwlLC}.  

%%%%%%%%%%%%%%%
%FVar
%%%%%%%%%%%%%%%

%%%% FIGURE 7 %%%%%%%%

\begin{figure}[ht]
\epsscale{1.15}
\plotone{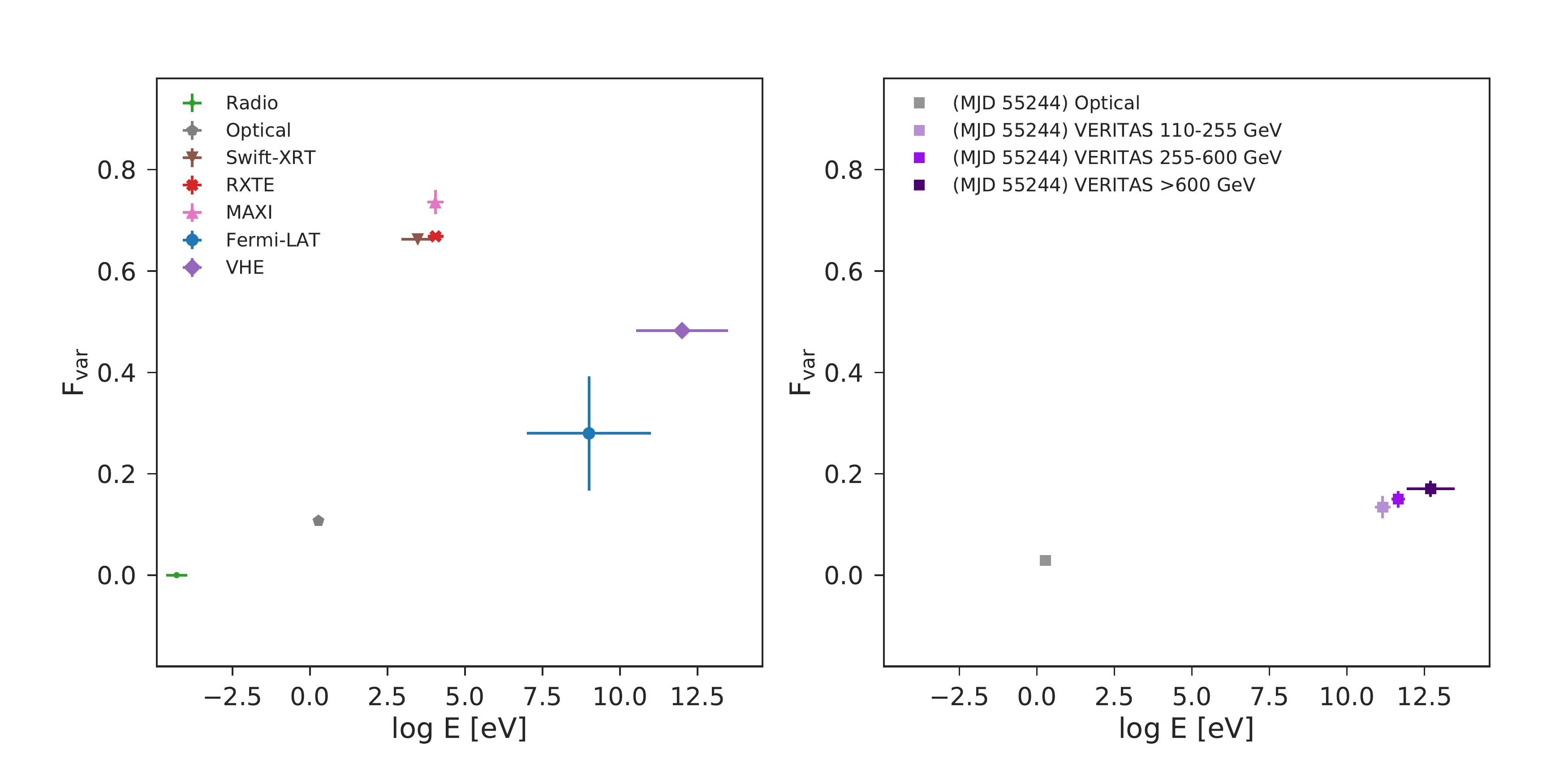}
\caption{\emph{Left:}  Fractional variability for each waveband over the full dataset shown in Figure~\ref{fig:mwlLC} (key in top left). The VHE band uses the nightly averaged binning from Figure~\ref{fig:mwlLC} for  both the VERITAS and the MAGIC light curves; the optical and radio bands include observations from all participating observatories displayed in Figure~\ref{fig:mwlLC}. \emph{Right:}  \emph{$F_{var}$} calculated for the 2-minute-binned light curves produced for the optical and three separate VERITAS energy bands on the night of the main flare as shown in Figure~\ref{fig:veropt_lcs} (key in top left).}
\label{fig:fvar}
\end{figure}

\subsection{Multiwavelength Variability}
\label{fvar}

We calculated the fractional root mean square variability amplitude, \emph{$F_{var}$}~\citep{edelson90,rodriguez97} -- as defined by Equation 10 in~\citet{vaughan03} with its uncertainty given by Equation 7 in~\citet{Poutanen08} -- for each available band, with the results shown in Figure~\ref{fig:fvar}. The \emph{$F_{var}$} calculation was performed  for the full duration of the light curves shown in Figure~\ref{fig:mwlLC} and, separately, for the 2-minute-binned optical and VERITAS (3 bands) light curves from the night of the giant flare (MJD 55244). Note that the four radio bands are shown under a single point covering the energy range of the bands, as no excess variance (\emph{$F_{var} = 0$}) was found in any of the bands.

The \emph{$F_{var}$} values from the full light curves spanning the month of February, 2010 increase from radio to optical to X-ray, drop again for the HE band, and then show maximal \emph{$F_{var}$} for the VHE band. This ``double-humped'' \emph{$F_{var}$} characterization, which has been observed in Mrk\,421 during low and high activity \citep[][]{Aleksic2015mid, Aleksic2015March2010, Balokovic2016}, could reflect the global difference in cooling time between the populations of electrons underlying the different bands. However, no strong conclusions can be drawn from these values as the integration times differ drastically for the light curves from different instruments potentially introducing large biases. 

The \emph{$F_{var}$} values for the optical and VERITAS light curves from MJD 55244 are more reliable for inter-band comparison, showing higher values for the VERITAS bands compared to the optical and an indication of an increasing trend with energy within the three VERITAS bands (though the \emph{p}-value of a $\chi^2$ difference test between a linear and constant fit is 0.13). If the particle-cooling timescale (with inverse-Compton scattering or SSC) is longer than the dynamical timescale of the emission region, the increasing \emph{$F_{var}$} with energy observed in the VHE can be related to the difference in cooling times between particles of different energies. The higher-energy particles will cool faster,  producing larger variability and a correspondingly higher \emph{$F_{var}$} value for a given timescale than lower-energy particles that cool more slowly.

There is a large contrast between the impressive flux variations at high energies and the muted behavior both in optical flux and linear polarization seen in Figure~\ref{fig:mwlLC}. 
The optical data show a smooth decrease of 20\% over the entire period. 
Two ``fast" variations (1-2 day timescales) of about 15-20\% are noted: one on MJD 55236 (8 February, 2010) in Epoch 1 in the ``preflare" time interval and the other in Epoch 2 on MJD 55244 (16 February, 2010), the night before the $\sim$11 CU (above 110 GeV) flare measured with VERITAS.
This latter fast optical variation is during the period where the HE and X-ray observations show some evidence for correlation (see Section \ref{HE-xray-Correlation}). 
It is interesting to note that, while the source clearly stayed high on 17 February, 2010 in X-rays and VHE, the optical flux diminished to values just slightly higher than the pre/post flare flux. 

 The optical polarization for Mrk\,421 increased from P=1.7\% to 3.5\% during 
 the VHE flare. No change in polarization position angle was detected over the same period, although larger
($\sim20^{\circ}$) position angle swings are observed just prior to and after the VHE flare. In general, both the variability in optical flux and polarization are mild during this period, with P=1--3.5\% and $\theta=125^{\circ}-155^{\circ}$. For comparison, the Steward Observatory monitoring data for Mrk\,421 obtained during the January, 2010 and March, 2010 observing campaigns show the blazar to be more highly polarized. For 14-17 January, 2010, P=3.7-5.0\%; $\theta=157^{\circ}-163^{\circ}$ and during 15-21 March, 2010, P=3.1-4.9\% with $\theta=114^{\circ}-130^{\circ}$. In addition, the object was about 0.3 mag brighter during the January, 2010 campaign compared to the February measurements, while it was $<0.1$ mag fainter in March, 2010. 

There are no signs of unusual activity in the radio observations of Mrk\,421 with the instruments that participated in this campaign (UMRAO, Mets{\"a}hovi and OVRO)  over the two weeks before and the two weeks after the main VHE flare. However, no observations were taken during the VHE flare night. High-resolution Very Long Baseline Array (VLBA) observations of Mrk\,421 were collected on 11 February, 2010 as part of the Monitoring Of Jets in Active galactic nuclei with VLBA Experiments (MOJAVE) program~\citep{Lister2018}. MOJAVE data on Mrk\,421 are also available from 17 December, 2009 and 12 July, 2010 observations. The 15~GHz MOJAVE images show significant extended structures associated with the source. The emergence of a potentially new component within the Mrk 421 \textit{mas} radio jet over the month following the giant flare was reported by \citet{Niinuma12} using the Japanese Very Long Baseline Interferometry (VLBI) Network, and also by~\citet{jorstad17} using observations from the VLBA Blazar program from Boston University. However, we cannot conclude that any of the components from MOJAVE or the VLBI observations are associated with the 17 February, 2010 VHE flare. The relative VLBA flux density, S$_{\text{VLBA}}$/S$_{\text{total}}$ (S$_{\text{total}}$ is the filled-aperture single-antenna flux density) from the 11 February, 2010 MOJAVE and 12 February, 2010 OVRO observations is comparable to the average historical value of $\sim 0.75$ from~\citet{Kovalev2005}. The parsec-scale jet direction reported in~\citet{jorstad17} is about $-25^{\circ}$ and the polarization angle of the radio knot B1 is about $-35^{\circ}$, both angles being approximately the same to those reported in Figure~\ref{fig:mwlLC} for the optical EVPA taking into account the ambiguity of EVPA with
respect to $\pi$.

%%%%%%%%%%%%%%%%%%%%%%%%%%%%%%%%%%%%%%%%%%%%%%%%%%%%%%%%%%%%%%%%%%%%%%
\subsection{VHE $\gamma$-Ray and X-ray Correlation Studies.} \label{VHE-xray-Correlation}

%%%%%%%%%%%%%%%%%%%%%%%%%%%%%%%%%%%%%%%%%%%%%%%%%%%%%%%

\begin{figure}[!htbp]
\centering
\epsscale{1.1}
\gridline{\fig{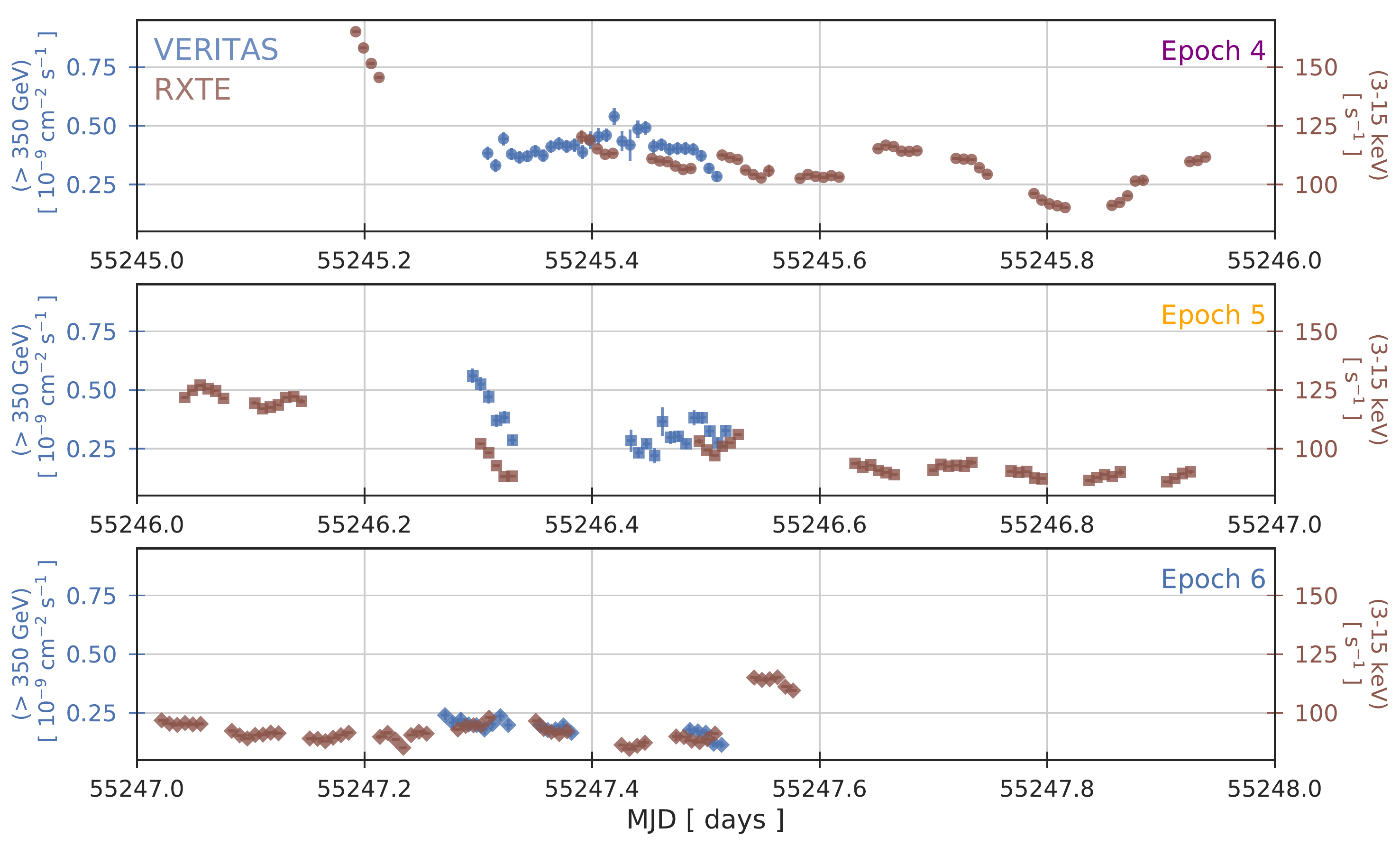}{1.\textwidth}{}}
\vspace{-3em}
\gridline{\fig{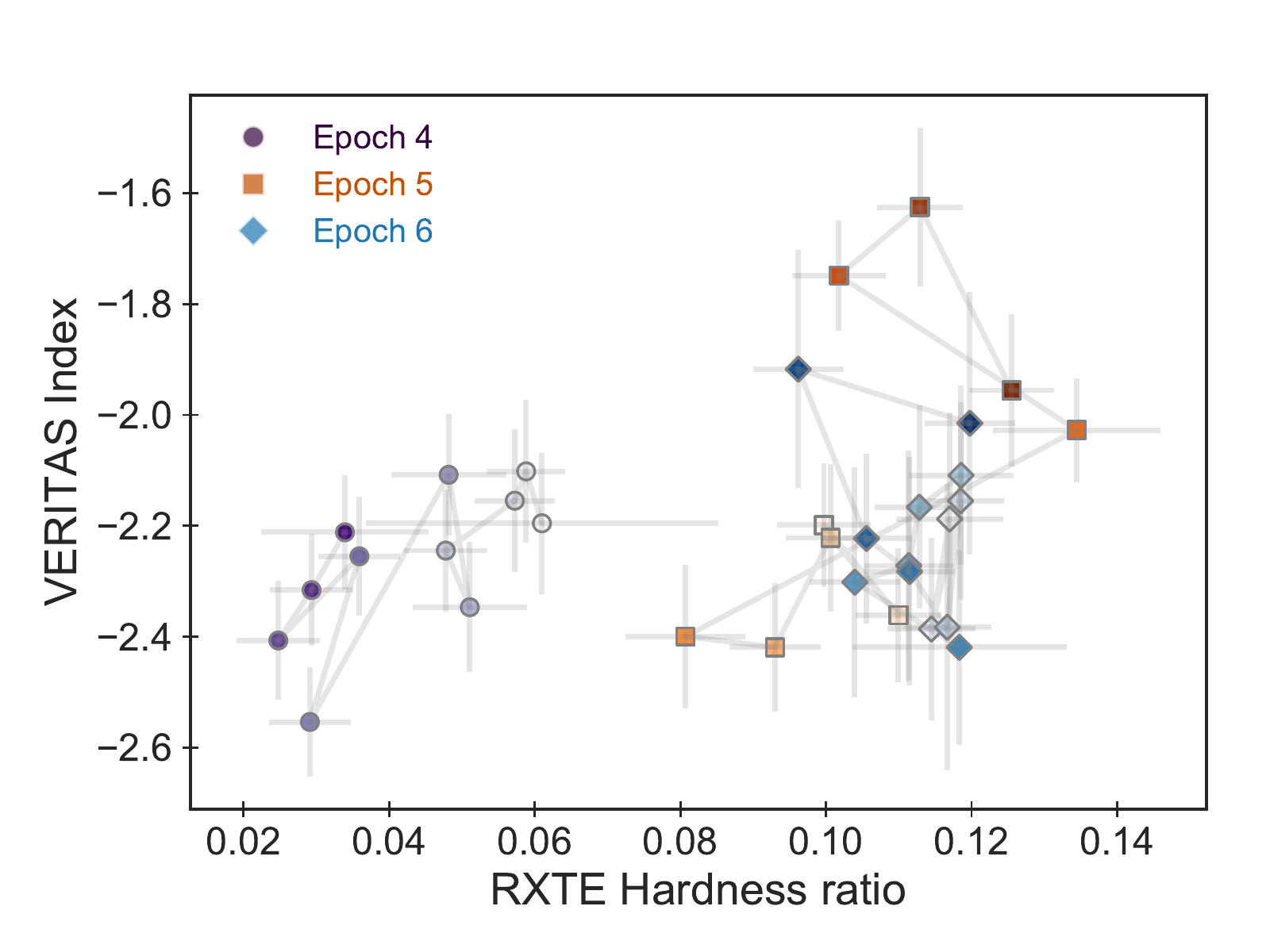}{0.49\textwidth}{}
          \fig{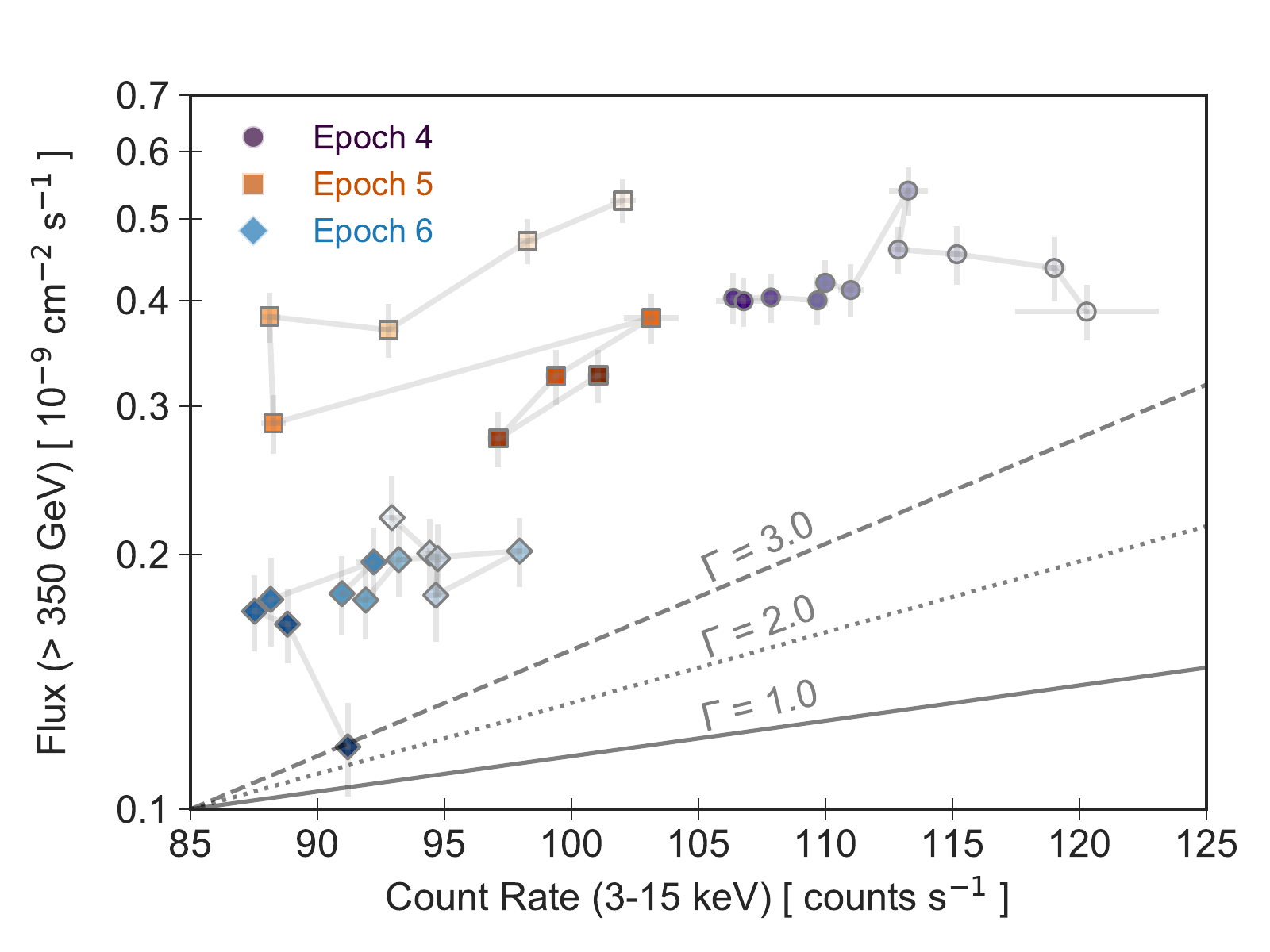}{0.49\textwidth}{}}
\vspace{-2em}
\caption{\textit{Top three panels:} The detailed 10-minute-binned light curves for Epochs 4-6 from Figure~\ref{fig:rxte_hysts_epochs} with the  VHE (VERITAS-blue) and the X-ray (\textit{RXTE}-brown) overplotted in the same panel to better highlight trends described in the text. From top to bottom, the panels are Epoch 4, Epoch 5 and Epoch 6.  \textit{Bottom two panels:}    VERITAS photon indices versus \textit{RXTE} hardness ratios (\textit{left}) and VERITAS and RXTE flux-flux correlation (\textit{right}) plots based on the 10-minute-binned light curves for each of the epochs in the above panels. Only points are plotted here that correspond to the ``matched" points in Figure~\ref{fig:xray-VHE-ilc} where there is strictly simultaneous overlap between \textit{RXTE} and VERITAS observations. Gray lines and color gradients are intended to guide the chronological progression of the points. The hardness ratio is taken between the 5-15 keV and 3-5 keV bands while the VHE indices are found from a power-law fit between 350 GeV and 3 TeV.}
\label{fig:ii-ff-xray-VHE}
\end{figure}

By visual inspection of Figure~\ref{fig:mwlLC}, we cannot ascertain whether the VHE flare was observed at its peak or on the decline. Furthermore, there is no overlapping \textit{Swift}-XRT or \textit{RXTE} data during the night of the highest VHE flux. Unfortunately we therefore cannot determine any correlation between X-ray and VHE at the peak observed in either band. There are only three \textit{Swift}-XRT exposures over the first two days of the decline, averaging 3.6 ks per exposure. \citet{Kapanadze2018} analyzed these data along with all available \textit{Swift}-XRT data for the period 2009-2012.  However, \textit{RXTE} data comprise eight short (average 3.6 ks) and five long (from 9.8 ks to 48.2 ks) observations during the decline Epochs 4-6 which overlap with VERITAS data. We thus use the \textit{RXTE} data for our in-depth VHE-X-ray studies; both these data sets are shown in Figure~\ref{fig:xray-VHE-ilc} with a zoomed-in version overplotting \textit{RXTE} and VERITAS data for Epochs 4-6 in the top three panels respectively of Figure~\ref{fig:ii-ff-xray-VHE}.  Clear inter-day variability is evident in both the VHE and X-ray bands with Epoch 4 mainly comprising a strong decay in an X-ray flare (40\% drop in PCA rate) while the VHE shows a slight rising trend, Epoch 5  catches the tail of another smaller X-ray flare followed by a rise - both mirrored in the VHE, and Epoch 6 shows minimal X-ray and VHE variability. 

%%%% FIGURE 8 %%%%%%%%

In Figure~\ref{fig:xray-VHE-ilc} we also show the VERITAS photon indices as well as the \textit{RXTE} hardness ratio in 10-minute time bins (lower two panels). The hardness ratio (HR) is taken between the 5-15 keV and 3-5 keV bands while the VHE indices are found from a power-law fit between 350 GeV and 3 TeV. Here we note that the overall trend for the X-ray data is an increase in the hardness ratio while the source in general is becoming steadily weaker in the X-rays (top panel of Figure~\ref{fig:xray-VHE-ilc}). On the other hand, the VHE data show no general trend during the decline phases of the flaring period. However, there are periods when the VHE indices become significantly harder than $\Gamma \sim -2$, indicating the VHE emission is in part below the IC peak frequency. 
Some of these exceptionally hard indices correspond to instances in which the VHE flux is at its weakest in this dataset. This is especially true towards the end of Epoch 6.  

The bottom left panel of Figure~\ref{fig:ii-ff-xray-VHE} looks at the VHE index versus X-ray hardness ratio over the full decline phase but restricted to data pairs where there is an exact time match between the VERITAS and \textit{RXTE} data (gray bands in Figure~\ref{fig:xray-VHE-ilc}). The data suggest a clustering around distinct states that represent ``snapshots" of the evolving system over several days. The cluster of extremely hard VHE indices and high X-ray HR values corresponds to a weak-flux state observed in both bands. Though unusual, these observations could indicate both the synchrotron and IC peaks  shifting together to higher frequencies  (without an increase in peak luminosity).  

As there are substantially more X-ray data than VHE data throughout the decline epochs, in Appendix~\ref{appendix_xray_hysteresis} we carry out a detailed examination of the X-ray data searching for evidence of hysteresis in the relationship between HR and the count rate. All epochs show a considerably different evolution of the hardness ratio with flux, with a variety of loops and trends exhibited even as  the overall \emph{increase} in HR is seen as the source \emph{weakens} in X-ray across the decline (as noted in Figure~\ref{fig:xray-VHE-ilc}). The standard harder-when-brighter scenario is only distinctly observed in Epoch 5.

%%%% TABLE 2 %%%%%%%%

\begin{deluxetable}{lcccc}[!htbp]
\tablecaption{Results from fits to the data for the left panel of Figure~\ref{fig:ii-ff-xray-VHE} with the relation $F_{\gamma} \propto F_X^{\Gamma}$, where $F_\gamma$ is the VERITAS flux above 350 GeV in units of 10$^{-9}$ cm$^{-2}$ s$^{-1}$, $F_X$ is the \emph{RXTE} count rate between 3~keV and 15~keV, and $\Gamma$ is the index. The Pearson's $\rho$ is shown along with the $p$-value for each fit.\label{tab:RXTEVerFits}}
\tablehead{Dataset & $\Gamma$ & $\chi^{2}$/NDF & $\rho$ & $p$-value }
\startdata
Full & $3.3 \pm 0.2$ & 220/30 & 0.76 & $3.6\times 10^{-7}$ \\
Epoch 4 (all) & $1.5 \pm 0.07$ & 16/8 & 0.22 & 0.52\\
Epoch 4 (first 4) & $ -1.6 \pm 0.18$ & 0.85/2 & -0.86 & 0.14 \\
Epoch 4 (first 5) & $ -3.2 \pm 1.2$ & 5.4/3 & -0.78 & 0.17\\
Epoch 4 (last 6) & $ 0.7 \pm 0.09$ & 0.34/4 & 0.57 & 0.24\\
Epoch 5 (all) & $2.0\pm 1.0$ & 37/7 & 0.35 & 0.36 \\
Epoch 5 (first 4) & $2.5 \pm 0.8$ & 1.5/2 & 0.92 & 0.078\\
Epoch 5 (last 5) & $1.6 \pm 1.0$ & 1.3/3 & 0.74 & 0.15\\
Epoch 6 (all) & $1.9 \pm 0.1$ & 6.9/11 & 0.48 & 0.095 \\
Epoch 6 (no last point) & $1.6 \pm 0.3$ & 2.1/10 & 0.64 & 0.024 \\
\enddata
\end{deluxetable}

To further investigate the flux-flux relationship between the  
synchrotron and IC peaks during Epochs 4-6, we show the VHE--X-ray flux-flux plot in the bottom right panel of Figure~\ref{fig:ii-ff-xray-VHE} for each epoch, where the X-ray and VHE data are simultaneous (indicated by the  gray bands in Figure~\ref{fig:xray-VHE-ilc}). We also show the linear, quadratic and cubic slopes corresponding to the relation $F_{\gamma} \propto F_X^{\Gamma}$ with fit values shown for each epoch displayed in Table~\ref{tab:RXTEVerFits}, along with the slopes for subsamples of the data in each epoch as well as the full dataset. For simple SSC behavior, we would expect to see correlation between the X-ray and VHE emission with a linear correlation slope indicating the system was in the Klein-Nishina (KN) regime \citep{Tavecchio1998}. In fact, the VHE-X-ray flux-flux plot shows inconsistent behavior across the three epochs. When considering the first four points, Epoch 4 shows a hint of an anti-correlation between the VHE and X-ray bands  which would be very inconsistent with a single-zone SSC model. Taking the last six points of Epoch 4, no correlation is seen -- the VHE stays roughly constant in flux as the X-ray dims. Epoch 5 captures a fast decrease in both VHE and X-ray, followed by a less dramatic rise in both bands. Both the fall and rise states show a correlation between the two bands, however with a $\sim$quadratic behavior in both ``cooling" and ``acceleration".   
Epoch 6 shows an erratic, uncorrelated relationship in time between the X-ray and VHE bands, though with a global fit nearly quadratic in slope.
Taken together, the range of behavior across the decline epochs between and within the X-ray and VHE bands is difficult to interpret as the evolution of the system in the context of a single-zone SSC model.

\subsection{HE $\gamma$-Ray and X-ray Correlation Studies.} \label{HE-xray-Correlation}
By inspection of the light curve in Figure~\ref{fig:mwlLC}, Epoch 2 shows an increase in both the MAXI X-ray and \textit{Fermi}-LAT HE $\gamma$-ray daily binned fluxes the day prior to the VHE flare observed with VERITAS. A simple test for variability was performed on the \textit{Fermi}-LAT light curve.
This yielded an improvement in log-likelihood over a constant model equivalent to $\chi^2=39.2$ for 23 degrees of freedom, corresponding to a p-value = 0.018. The MAXI light curve is clearly variable ($\chi^2/\mbox{NDF} = 930/23$; p-value$\sim$0).

A preliminary cross-correlation analysis using \textit{Fermi}-LAT ``Pass7'' P7SOURCE\_V6 event selection and instrument response functions  found that the lag between the X-ray and HE $\gamma$-ray light curves was consistent with 0 days \citep{Madejski2012}. 
We performed an analysis using the ``Pass8'' \textit{Fermi}-LAT data and IRFs  corresponding to those used to generate the \textit{Fermi}-LAT light curve in Figure~\ref{fig:mwlLC}. A linear correlation coefficient was  calculated for the time-matched MAXI and \textit{Fermi}-LAT fluxes, resulting in a mean value of $\rho = 0.54\pm0.12$. The mean value and the $1\sigma$ uncertainties of the linear correlation coefficient were determined by resampling both light curves within measurement uncertainties over 100,000 iterations. 

\begin{figure}
\epsscale{1.15}
\plotone{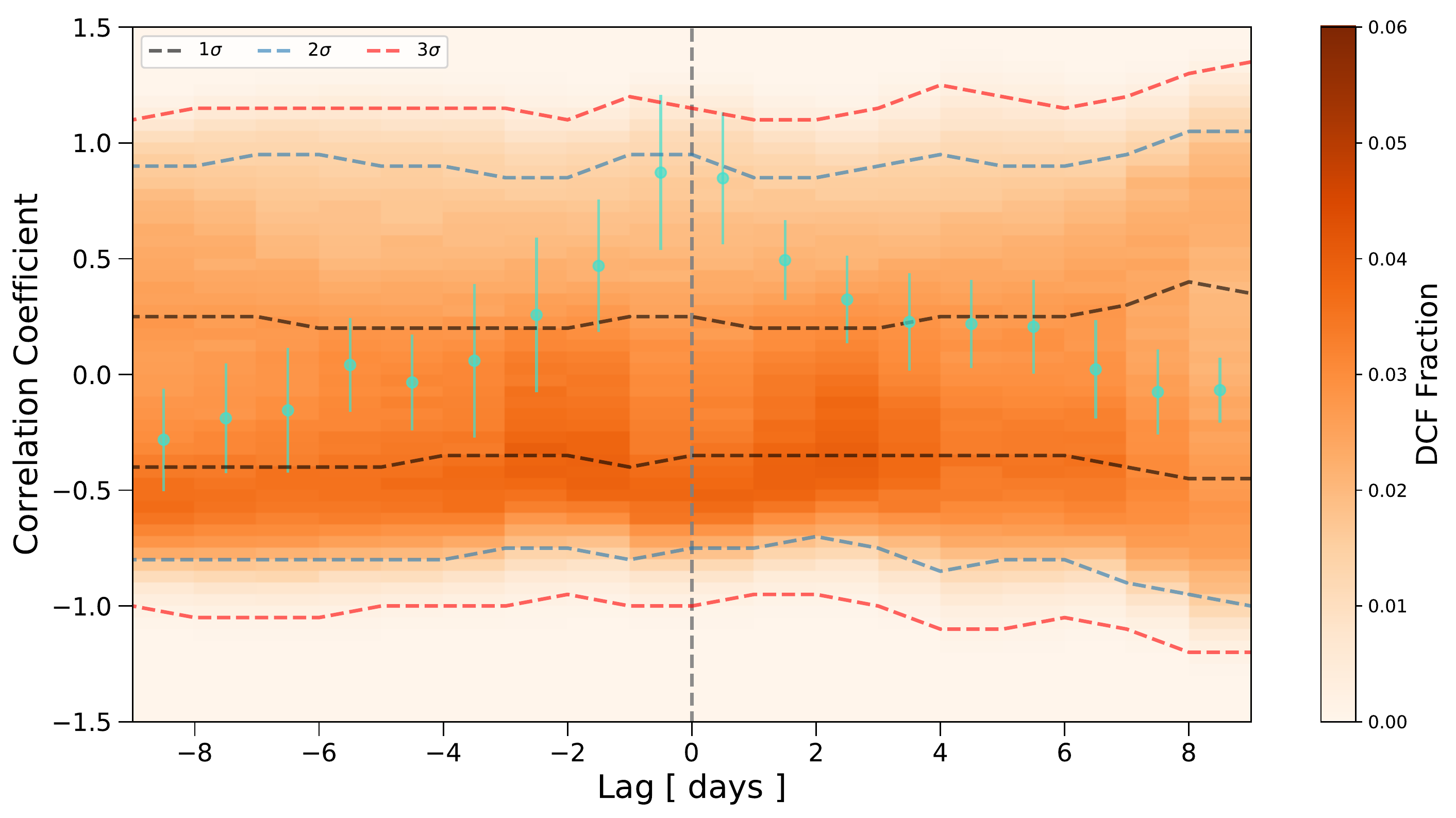}
\caption{The DCCF calculated from the observed MAXI--\textit{Fermi}-LAT light curves is shown with turquoise points. Correlation significance levels (shown with dashed lines) are estimated through a Monte Carlo method. 
During each iteration, the observed \textit{Fermi}-LAT light curve is paired with a light curve simulated from the MAXI PSD to calculate a simulated DCCF.
``DCF fraction'' represents the fraction of times a simulated DCCF falls in a given lag-time and correlation-coefficient bin (shown with the 2D histogram color map). The DCF fraction histogram (representing a PDF) is integrated to obtain the confidence levels.
\label{fig:dcf_fermi_maxi}}
\end{figure}

To further investigate this potential correlation, we conducted a DCCF analysis between MAXI--\textit{Fermi}-LAT light curves in the manner described in Section~\ref{OpticalCorrelation}.  In this case, the PSD from the MAXI light curve was fit using the method by~\citet{Max-Moerbeck2014} and the best-fit MAXI PSD ($P(f) \propto f^{-1.95}$) was used to generate 100,000 random light curves paired with observed \textit{Fermi}-LAT light curve (the conservative approach), with the results shown in Figure~\ref{fig:dcf_fermi_maxi}.

We find a $\sim 2\sigma$ correlation at a lag of $\sim 0$ days. 
The confidence level of the correlation at $\sim 0$ days is considerably higher ($\sim 4\sigma$) if light curves are simulated from PSDs for \textit{Fermi}-LAT as well (with best-fit PSD, $P(f) \propto f^{-1.75}$). The PSD fit errors are very large, however, making it difficult to characterize the uncertainties on the significance of the correlation. 

%%%% FIGURE 9 %%%%%%%%

%%%%%%%%%%%%%%%%%%%%%%%%%%%%%%%%%%%%%%%%%%%%%%%%%%%%%%%%%%%%%%%%%%%%%%
%%%%%%%%%%%%%%%%%%%%%%%%%%%%%%%%%%%%%%%%%%%%%%%%%%%%%%%%%%%%%%%%%%%%%%
%%%%%%%%%%%%%%%%%%%%%%%%%%%%%%%%%%%%%%%%%%%%%%%%%%%%%%%%%%%%%%%%%%%%%%
\section{Discussion and Conclusions} \label{conclusions}

The VHE flare observed from Mrk\,421 in February, 2010 is an historically significant flare. During the night of the giant flare observed with VERITAS, Mrk\,421 reached a peak flux of about 27 CU Units above 1~TeV. 
This episode rivals the brightest flares observed from any source in VHE $\gamma$-rays, including the extraordinary flare of PKS 2155-204 in 2006 detected by H.E.S.S.~\citep{Aharonian2009a} and the 27 February, 2001 flare of Mrk\,421 seen with Whipple~\citep{Krennrich01,ACCIARI2014}. Another exceptionally strong flare in Mrk\,421 was detected by both VERITAS and MAGIC in April, 2013 \citep{2013ATel}. 
As extreme as the currently reported flare is, it is unclear from the analyses described in this paper and summarized below whether this represents a fundamentally different behavior state for this object or just an extreme end of the same underlying processes that have yielded the range of behavior previously reported.

\subsection{VHE-Optical Band Correlation}
A  cross-correlation analysis was performed between the VHE and optical bands during the night of the VHE flare. 
The observed optical and VERITAS 2-minute-binned light curves exhibit a $3\sigma-4\sigma$ significance correlation with an optical lag of 25-55 minutes centered at 45 minutes.
Such behavior can be accommodated under a single-zone SSC scenario, in which the emission in both VHE and optical bands is produced by a single distribution of electrons. Under this scenario, the optical lag could be explained by the slower cooling of the less energetic electrons that underlie the optical data compared to the electrons responsible for the VHE emission~\citep{Boettcher2019}. The lag timescales can be used to set an additional constraint on the magnetic field strength for future SED modeling efforts.

VHE-optical correlation has been previously observed in HBLs, but not with a lag and not at the short timescales probed by this unprecedented dataset. For example,  the 2008 multiwavelength campaign on PKS 2155-304 reports a $> 3\sigma$ correlation between the H.E.S.S. data and the $V$, $B$, and $R$ optical bands  on daily timescales and with no lag \citep{Aharonian2009a}. It is more common to observe a correlation between the HE and optical which is likely explained by both bands arising from the same electron population in the simple SSC model \citep{Cohen2014}. 

\subsection{Fast Variability in VHE $\gamma$-rays}

The exceptional brightness of the flare on 17 February, 2010 in VHE enabled VERITAS to produce 2-minute-binned light curves with $10\sigma$ significance in each bin yielding strongly-detected, short-term variability. 
The variable emission within the first 95 minutes of VERITAS observations on that night can be described by at least two successive bursts. 
\textit{Burst 2} is characterized by an asymmetric profile with a faster rise time, followed by a slower decay. This behavior has been previously observed~\citep[e.g.,][]{Zhu2018} and is typically attributed to emission from electrons with longer cooling than the dynamical timescales, assuming both the cooling and dynamical timescales are much longer than the acceleration timescale. Under this scenario, the flare rise time is related to the size of the emission region~\citep[e.g.,][]{Zhang2002}.

Assuming the above conditions, we used the rise timescale of \textit{Burst 2} to place an upper limit on the size of the emission region associated with the burst, $R_{B}$:

\begin{equation}
R_{B} \leq \frac{c t_{\text{var}} \delta}{1 + z},
\end{equation}
where $c$ is the speed of light, $t_{var}$ is the variability timescale, and $\delta$ is the Doppler factor. Using the most likely \textit{Burst 2} rise time of 22 minutes for $t_{\text{var}}$, we obtained $\delta^{-1}R_{B} \lesssim 3.8\times10^{13}$~cm.

Furthermore, the time variability of the VHE flux, in conjunction with compactness and opacity requirements of the emitting region, can be used to give an estimate of the minimum Doppler factor of the ejected plasma in the jet of the blazar. Following \citet{Tavecchio1998} and \citet{Dondi95}, the minimum Doppler factor was calculated using:

\begin{equation}
\delta_{\text{min}} > \Bigg[\frac{\sigma_{\text{T}}}{5hc^2}d_L(1+z)^{2\beta}\frac{F(\nu_{0})}{t_{\text{var}}}\Bigg]^{1/(4+2\beta)},
\end{equation}
where $\sigma_{\text{T}}$ is the Thomson scattering cross-section, $d_{L}$ is the luminosity distance of the source, $z$ is the redshift, $t_{\text{var}}$ is the observed variability timescale, and $F(\nu_{0})$ and $\beta$ are the flux and spectral index, respectively, of the target photons of the $\gamma$-rays for pair production.

To estimate the Doppler factor limit, we used the following parameters: the observed variability timescale $t_{\text{var},\,\text{VHE}} = 22$~minutes;  the $\gamma$-ray photon energy $E_\mathrm{\gamma} = 110$~GeV, corresponding to a target photon frequency of $6.0\times10^{14}$~Hz (500~nm) for maximum pair-production cross section; and the spectral index, $\beta=-0.16$, and $F(\nu_{0})=1.35$~mJy of the low-energy photons derived from the 3 \textit{Swift}-UVOT-band observations during MJD 55244-55246. The latter value was obtained using, $F(\nu_{0}) = F_{uvw1}\big(\nu_{uvw1}/\nu_{0}\big)^{\beta}$.  
Assuming these parameters, the derived Doppler factor lower bound is $\delta_\mathrm{min}\gtrsim 33$.
The fast variability measured with this dataset results in a larger Doppler factor compared to \cite{Blazejowski2005}, where a lower limit on the Doppler factor of $\delta_\mathrm{min}\gtrsim 10$ was obtained with an $\sim$hour-scale time variability in the VHE data from the  3-CU flare of Mrk\,421 during April, 2004.

For the overall system to be consistent with reported lower Doppler factors from VLBI measurements, results from fast flares such as that reported here indicate that the $\gamma$-ray emission zone may be smaller than the jet-cross section. 
For example, \citet{Giannios2013} suggests that rapid $\sim$minute-scale flares on an ``envelope'' of day-scale flares can be due to large plasmoids created during a magnetic reconnection event. However, \cite{Morris2018} show that while such a ``merging plasmoid'' model can explain the VHE light curve from the 2016 fast flare from BL Lac \citep{Abeysekara2018}, it has difficulty reproducing the SED. 

A potential counter-clockwise loop (known as spectral hysteresis), or a harder-when-weaker trend, is present in the index versus flux representation for \textit{Burst 1}, while the photon index is essentially constant for \textit{Burst 2} even as the flux changes by a factor of $\sim$1.5. Spectral hysteresis can occur as a result of competing acceleration, cooling, and dynamical timescales, which determine how the effects of particle injection into an emitting region translate to the observed photons~\citep{kirk98,LiKusunose2000,bottcher02}. Counter-clockwise hysteresis is related to a case in which dynamical, acceleration, and cooling timescales are comparable. The change in the number of emitting particles in this scenario is determined by the acceleration process, which proceeds from lower to higher energies and leads to higher-energy photons lagging behind the lower-energy photons.

A modified auto-correlation analysis is applied to the VERITAS data on the night of the flare to look for potential variability on short timescales; however, no significant time structures are found within 10--60~s timescales. Combining this result with timescales probed by the VHE PSD analysis, we conclude that the VHE emission is consistent with a pink noise characterization over a wide range of timescales -- from $\sim$seconds to $\sim$hours. Power-law PSDs in blazars have been detected in X-ray as well as VHE and are indicative of an underlying stochastic process \citep{Aharonian2007-exceptional}. A power-law PSD could also point to a self-organizing criticality (SOC) system, such as magnetic reconnection, as the underlying physical process responsible for the flaring behavior observed for Mrk\,421 \citep{Lu1991, Aschwanden2011, Kastendieck2011}. A recent study of Mrk\,421 flares extracted from archival \emph{XMM-Newton} X-ray data spanning from 2000 to 2017 is consistent with the expectations for a SOC model, thus lending support to the magnetic reconnection process driving blazar flares \citep{yan2018}. Additionally, the flatness of the PSD indicates that the turn-on/turn-off timescale of mini-flares can be below an hour and generally has a wide probability distribution extending from sub-hour timescales to entire nights \citep{Chen2016}.

\subsection{Multiwavelength Correlation Studies}
In addition to the optical-VHE correlation study, several other intra-band and multi-band correlation studies were carried out. The decay of the flare in the VHE and X-ray bands occurs over the course of four days.
Correlation studies between VHE (VERITAS) and X-ray (\emph{RXTE}) bands show a diverse and inconsistent range of behavior across the decline epochs. The flux-flux relationship between the synchrotron peak (as probed by the X-ray data) and the IC peak (as probed by the VHE data) moves in Epoch 4 from an indication of anti-correlation to no correlation. \cite{Blazejowski2005} report a lack of correlation seen in day-scale coincident VHE (Whipple) and X-ray (\emph{RXTE}) data, which is potentially explained by an X-ray flare leading the VHE flare by 1.5 days. The dataset reported in our work indicates lack of correlation between X-ray and VHE on the $\sim$10-minute timescales probing potentially quite different mechanisms.   To our knowledge, an anti-correlation between the X-ray and VHE has never before been reported for Mrk\,421.  Epoch 5 shows a $\sim$quadratic behavior in $F_{\gamma} \propto F_X^{\Gamma}$  most notably in the ``cooling'' segment of the epoch. This behavior  has been seen before in both Mrk\,421 \citep{Fossati2008} as well as in the exceptional flare in PKS 2155-304 \citep{Aharonian2009a} and is not consistent with the linear relationship expected from a system scattering in the Klein-Nishina regime \citep{Aharonian2009a}. However, Thomson scattering into VHE photon energies requires unacceptably large Doppler factors \citep{Katarzynski2005}.

The \textit{RXTE} results indicate spectral hardening as the source becomes fainter over this period. 
Such behavior can be an indication of the synchrotron peak shifting to higher frequencies as the flare decays, which would be unusual, or the possibility that the synchrotron photons in the keV band soften first, uncovering a population of harder photons produced in the keV band by the IC process at the beginning of the flare \citep{LiKusunose2000}.  On the other hand, no clear long-term trends are apparent in the VHE photon index as the flare decays.  Nonetheless, it is interesting to note that the VHE indices become harder than $-2$ at times during the decay period, indicating the Compton peak moves into the TeV regime even as the overall VHE flux is decreasing. The fact that both the X-ray and VHE data show a harder-when-weaker trend at the same time may indicate that both peaks have shifted and the source has temporarily become an extreme HBL~\citep{Costamante2001,Bonnoli2015,Cerruti2015}. 
 Time-dependent extreme HBL behavior has recently been reported for Mrk\,501 \citep{Ahnen2018} though changing on yearly timescales.
 
A correlation between HE and X-ray (MAXI) bands was observed on daily timescales. We found a $\sim 2\sigma$ correlation at a lag of $\sim 0$ days while a less conservative approach yielded $\sim 4\sigma$. While unusual, HE and X-ray correlations have been seen in other jetted systems including NGC 1275 and can indicate, for example, a fresh injection of electrons into the emission region \citep{Fukazawa2018}. 

\vspace{-0.2cm}
\subsection{Multiwavelength Variability}
A study of the energy dependence of the fractional variability ($F_{var}$) across all participating instruments resulted in a ``double-humped" structure that seems to be characteristic for Mrk\,421 in both flaring and quiescent states \citep{Aleksic2015mid,Aleksic2015March2010,Balokovic2016}. However, this is quite different from the $F_{var}$ characterization seen in the other well-studied nearby HBL, Mrk\,501, where a general increase in variability as a function of energy has been observed \citep{Aleksic2015Mrk501,Ahnen2017Mrk501, Ahnen2018}. While a strict comparison is difficult due to the vastly different integration times for the participating instruments in each campaign, the different $F_{var}$  dependence on energy between the two sources is likely attributed to the difference in the $F_{var}$ values in the X-ray band, with lower X-ray $F_{var}$ values typically seen in Mrk\,501. This could indicate that the X-ray instruments more often probe the rising edge of the synchrotron peak for Mrk\,501 than for Mrk\,421 which would be consistent with the synchrotron peak excursions to more extreme HBL regimes seen in Mrk\,501 \citep{Nieppola06,Ahnen2018}. The upcoming work studying the SEDs constructed from these data can further elucidate these observations.

\software{VEGAS~\citep{Cogan2008}, 
Event Display~\citep{Daniel2008},  MARS~\citep{Moralejo2009},
Fermi Science Tools\footnote{\url{http://fermi.gsfc.nasa.gov/ssc/data/analysis/documentation/Cicerone/}}, HEASoft (FTOOLS+XANADU)~\citep{heasoft2014}, REX\footnote{\url{http://heasarc.gsfc.nasa.gov/docs/xte/recipes/rex.html})}, Xspec~\citep{xspec1996}, emcee~\citep{emcee2013}} 

\clearpage
\appendix
\section{Power Spectral Densities}\label{sec:psds}
\vspace{-0.2cm}

%%%% FIGURE 10 %%%%%%%%
\begin{figure}[ht]
\epsscale{1.1}
\plotone{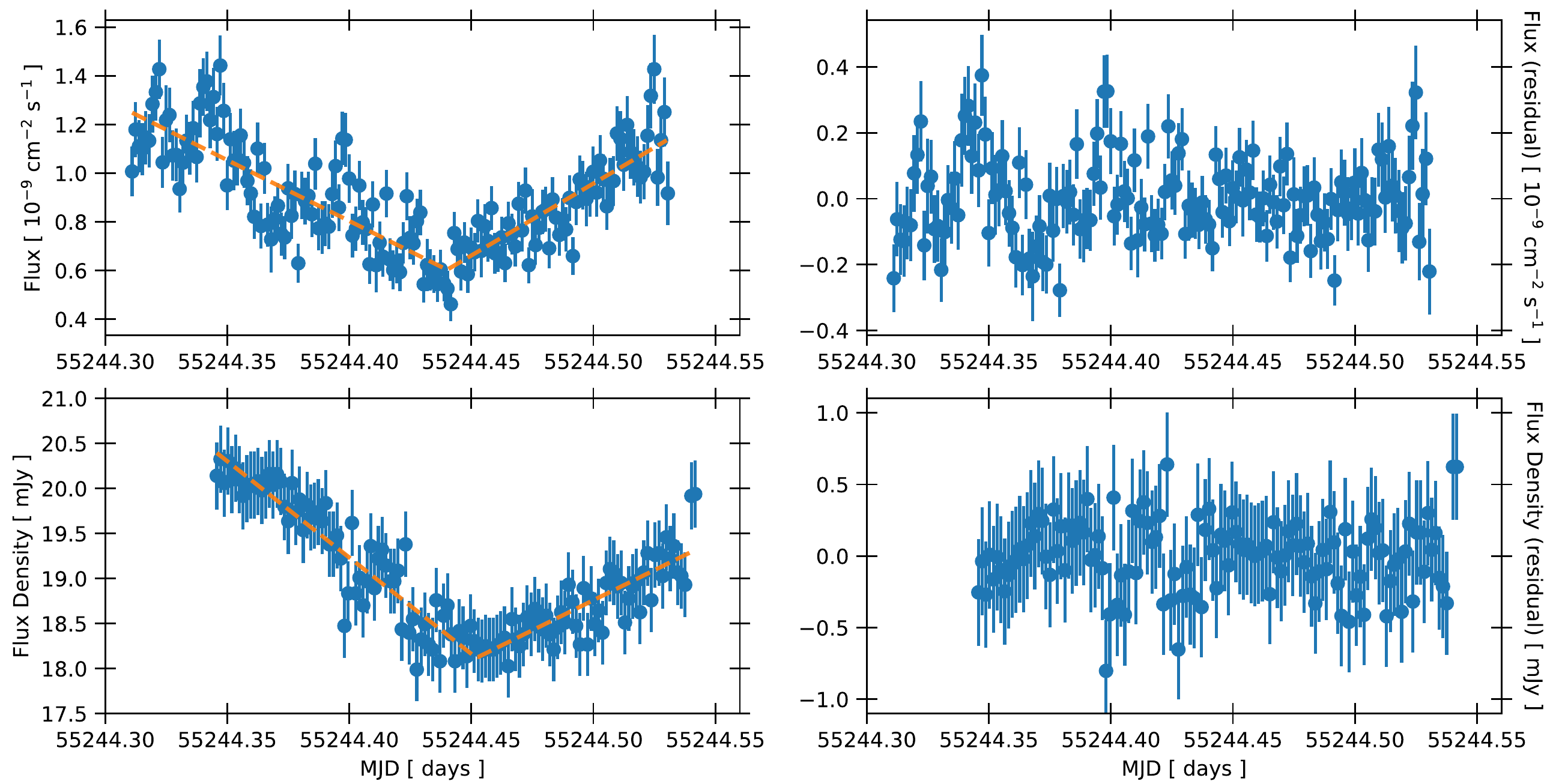}
\caption{2-minute-binned VERITAS $> 420$~GeV (\textit{top}) and optical R-band (\textit{bottom}) light curves. \textit{Left} panels are the observed light curves while \textit{Right} panels show the light curves after long-term trend removal.
\label{fig:veropt_detr}}
\end{figure} 

For the Mrk\,421 VERITAS and optical light curves, Power Spectral Densities (PSDs) were calculated using the FFT method available through the \textit{POWSPEC} program within the \textbf{XANADU} X-ray astronomical spectral, timing, and imaging software package.\footnote{https://heasarc.gsfc.nasa.gov/xanadu/xanadu.html}
PSDs were calculated with both the observed VERITAS and optical light curves as well as for those where the long-term trends have been modeled and removed. Trend removal was done to avoid potential contamination of higher-frequency signal by lower frequencies. A piecewise continuous linear function -- represented by a linear spline with a single node at the best-fit location -- was used to model and subtract the long-term trend in each light curve. The observed and detrended light curves are presented in Figure~\ref{fig:veropt_detr}. 
The entirety of both the VERITAS and optical light curves was used. Light curves are split into intervals within which the power spectra were independently calculated and later averaged. 

\begin{figure}[ht]
\epsscale{1.15}
\plotone{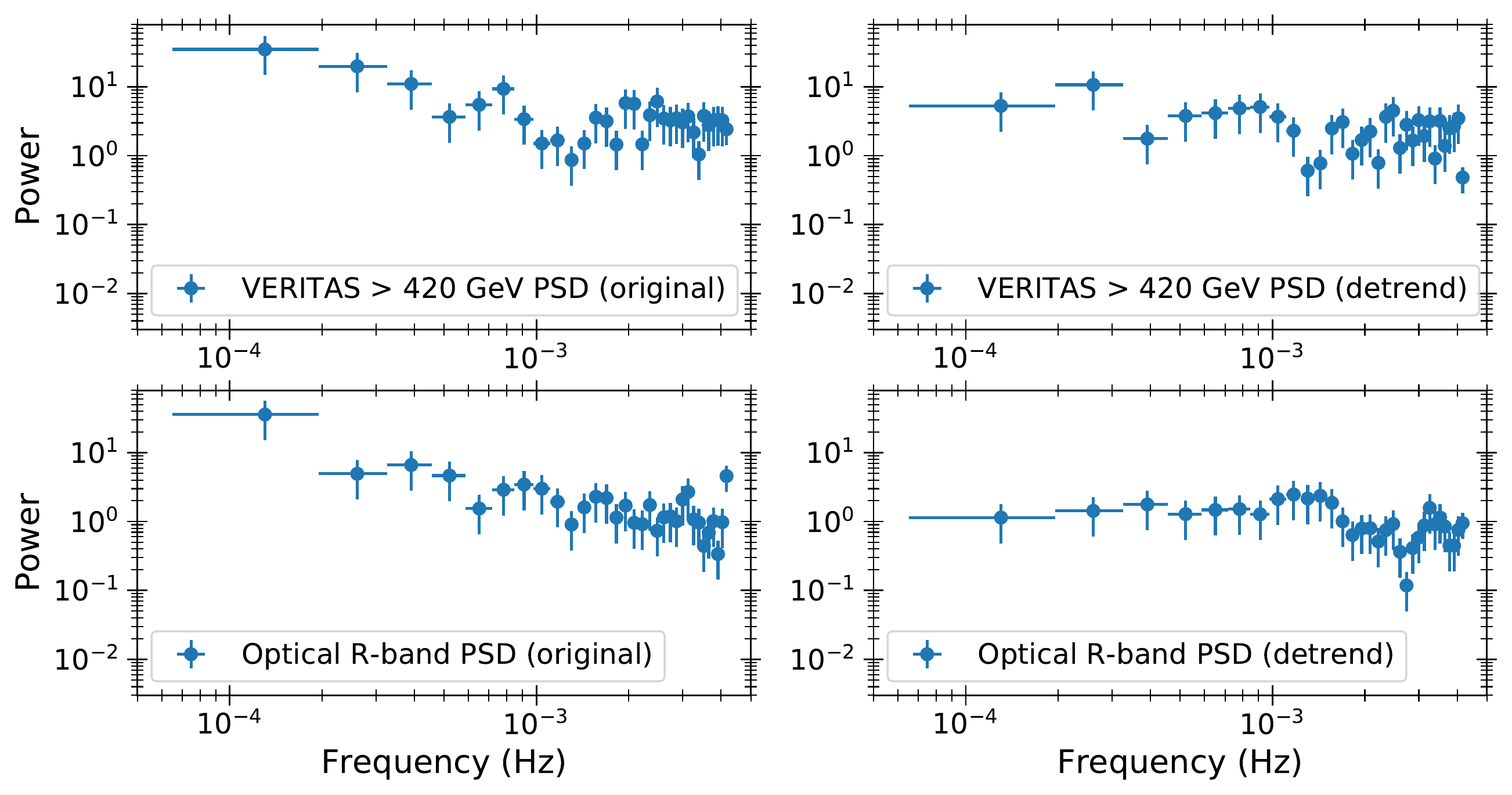}
\caption{VERITAS $> 420$~GeV (\textit{top}) and optical R-band (\textit{bottom}) PSDs with the original observed light curves (\textit{left}) and with light curves after long-term trend removal (\textit{right}).
\label{fig:veropt_psds}}
\end{figure}

The uncertainties on power in individual frequency bins were calculated as the standard deviation of the average of the power from different intervals. The resulting VERITAS and optical PSDs with and without detrending are displayed in Figure~\ref{fig:veropt_psds}. 
The best-fit power-law spectral indices for the VERITAS and optical PSDs were estimated with the method by~\citet{Max-Moerbeck2014} resulting in values of $-1.75$ and $-1.85$, respectively.
The detrended light curves were used to determine whether correlation existed at higher frequencies between the VHE and optical bands; no significant correlation was observed at least at short timescales.

%%%% FIGURE 11 %%%%%%%%

\section{Modified Auto-Correlation Analysis}\label{appendix_MACF}
The modified cross-correlation function ($MCCF$) is defined as a function of the lag time $\tau$, with $\tau = k\Delta t$:

\begin{equation}
    MCCF(\tau) = \sum_{i} \frac{(x_{1}(i\Delta t) - \bar{x_{1}}) (x_{2}(i\Delta t + \tau) - \bar{x_{2}})}{\sigma_{1}\sigma_{2}} ,
\end{equation}
where $x(t)$ is the number of counts in time bin $(t, t + \Delta t)$, and $\sigma$ is the standard deviation of $x$. For this modified correlation function, however, $\tau$ is not constrained to be an integer multiple of the light curve bin size, $\Delta t$, and can be incremented by the time resolution element, $\delta t$. The $MCCF$ can then be calculated for lag times, $\tau = m\,\delta t$ (with $m = 0, \pm 1,\pm 2, \ldots$) for light curves with a given timescale $\Delta t$. The timescale corresponding to the maximum of $MCCF(k\,\delta t) / MCCF(0)$ gives the lag time between $x_{1}$ and $x_{2}$.

From the definition of $MCCF$, the modified autocorrelation function ($MACF$) is obtained by setting $x_{1} = x_{2}$,

\begin{equation}
    MACF(\tau) = \sum_{i} \frac{(x(i\Delta t) - \bar{x}) (x(i\Delta t + \tau) - \bar{x})} {\sigma^{2}}.
\end{equation}
The FWHM (full-width at half maximum) of the $MACF$ is a measurement of the variability duration. The maximum of FWHM$_{MACF}/\Delta t$ may be treated as a characteristic timescale for the time series.

The MACF has the advantage over the regular ACF of being sensitive to variations on time scales smaller than the typical time scales within which significant excess is detected, potentially reaching the time resolution of the instrument $\delta_{t,\,\mathrm{VER}}$. Thus the MACF method does not use an ad-hoc time binning (for example the 2-minute bins from Figure~\ref{fig:veropt_detr}). Rather, the MACF of the VERITAS flux was constructed assuming a time resolution based on the minimum trigger rate of VERITAS during the Mrk\,421 flare observations, $\delta_{t,\, \mathrm{VER}} \sim 0.7$~ms. Then, for a range of $\Delta T$'s, where  $\Delta T$ is some integral number of $\delta_{t,\,\mathrm{VER}}$, the correlation was calculated between the initial time series and a new series shifted by the time resolution $\delta_{t,\,\mathrm{VER}}$. This ``sliding window" process allows the MACF to find any characteristic variability on timescales shorter than the time step $\Delta T$. 

\begin{figure}[h]
\epsscale{1.15}
\plotone{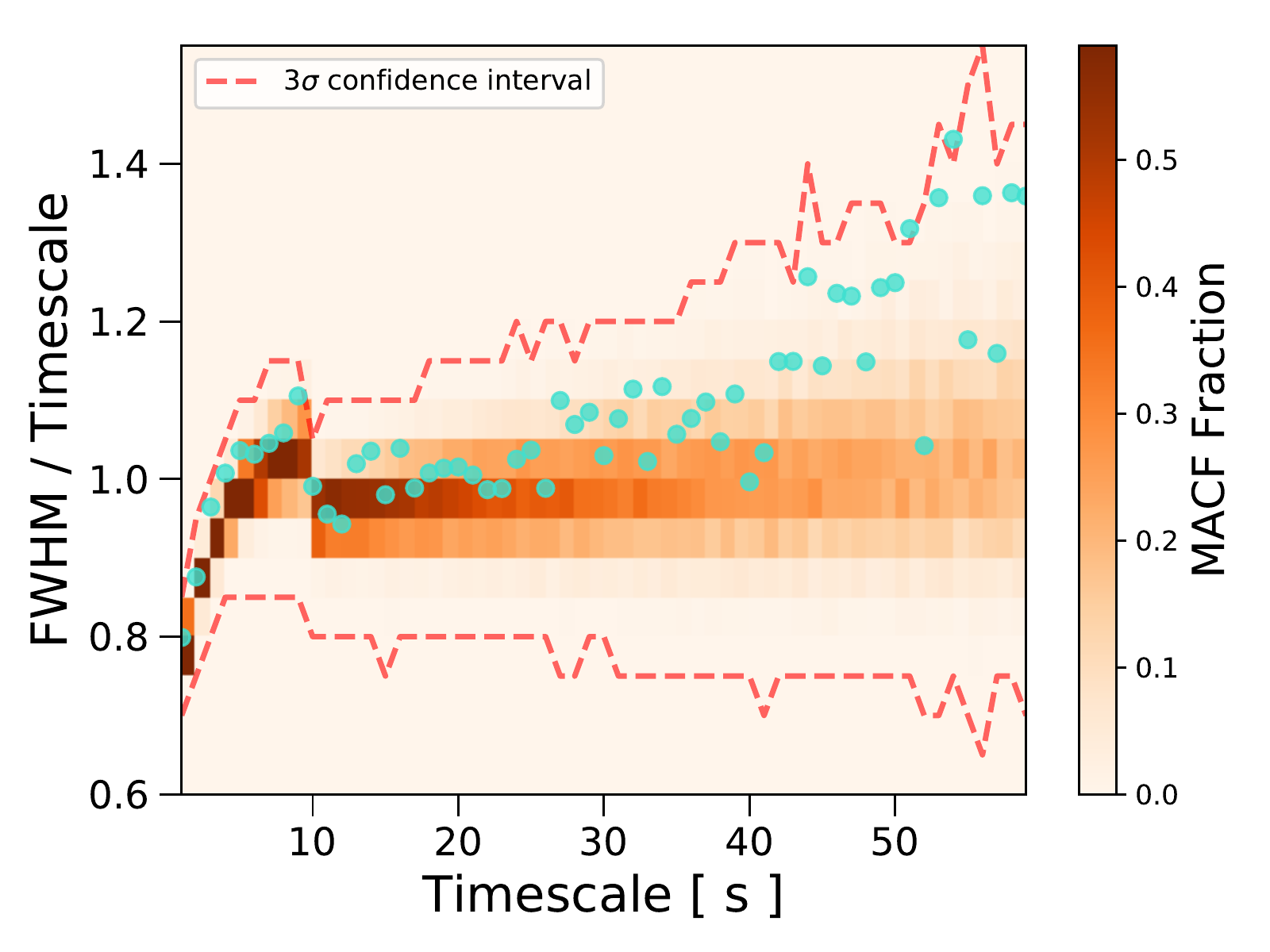}
\caption{Modified autocorrelation function for the VERITAS events above 420~GeV. Purple points show the MACF from VERITAS observations, while the color map bounded by red curves shows the region encompassed by MACFs from events simulated from a pink noise process.\label{fig:macf}}
\end{figure}

Here we have applied the MACF to the night of the VHE flare (Epoch 3; see Section~\ref{LightCurves}) using all events above an energy threshold $E=420$~GeV.
Figure~\ref{fig:macf} shows the FWHM$_{MACF}/\Delta T$ as a function of timescale where any characteristic timing signature would show up as a well-defined peak in the data points. MACFs from simulated events following a pink noise process (with $f^{-1.75}$ as above) were also calculated for 1000 iterations and are shown with the color map; they are bounded by dashed red lines at the 99\% confidence level.

%%%% FIGURE 12 %%%%%%%%

\section{X-ray Hysteresis Search}\label{appendix_xray_hysteresis}

\begin{figure}[!htbp]
\gridline{\fig{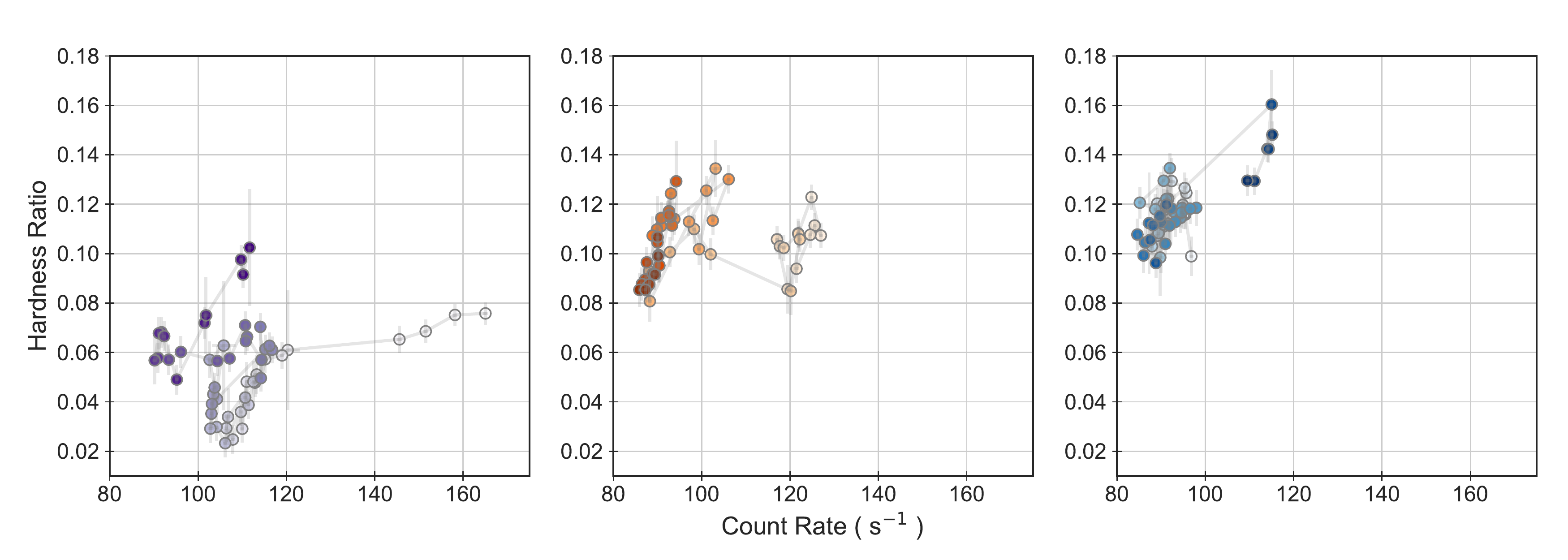}{1.\textwidth}{}}
\vspace{-3em}
\gridline{\fig{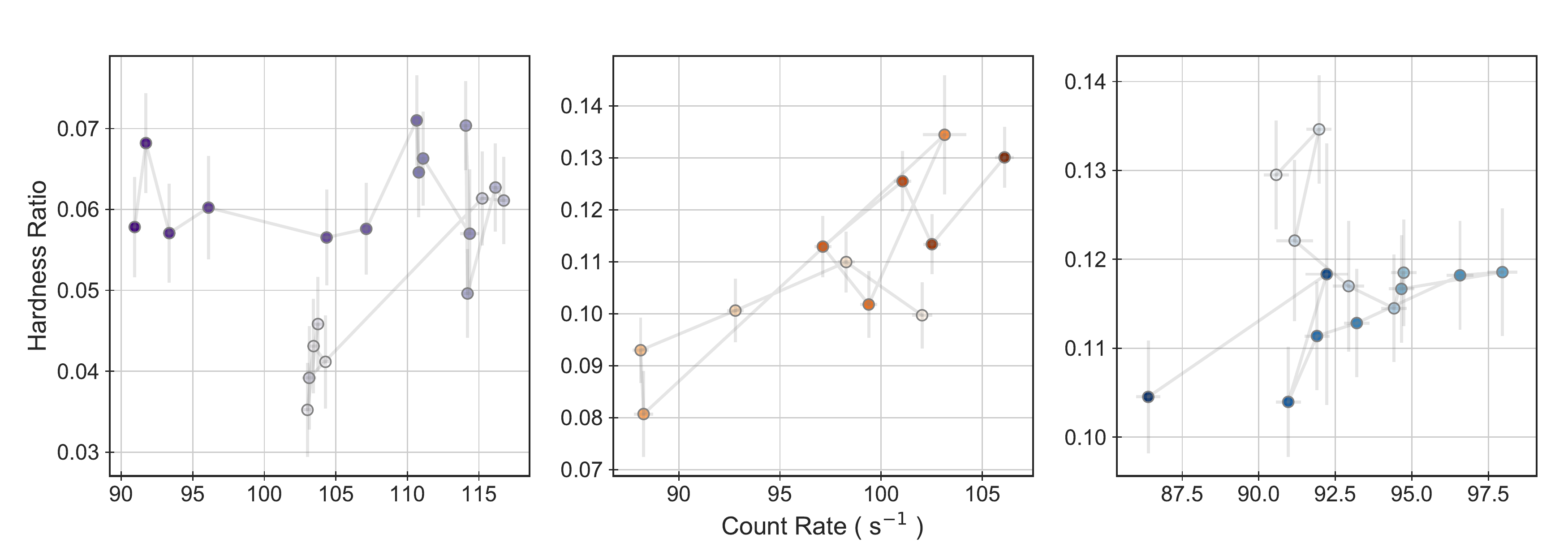}{1.\textwidth}{}}
 \caption{\textit{Top panels}: Hardness ratio (between 5-15~keV and 3-5~keV) versus count rate of the \textit{RXTE} observations of Mrk\,421 over the three decline epochs indicated by the arrows in Figure \ref{fig:xray-VHE-ilc} (not strictly simultaneous with VHE data within the epoch). (a) Figure on left: Epoch 4; (b) figure in middle: Epoch 5 ; (c) figure on right: Epoch 6. The color shades of the data points represent the chronological progression of the bursts, with lighter colors corresponding to earlier times. \textit{Bottom panels}: Zoom in on hardness ratio versus count rate of the \textit{RXTE} observations of Mrk\,421 corresponding to the colored bands in Figure \ref{fig:xray-VHE-ilc} (not strictly simultaneous with VHE data within the epoch). (a) Figure on left: Epoch 4 zoom-in; (b) figure in middle: Epoch 5 zoom-in; (c) figure on right: Epoch 6 zoom-in. The color shades of the data points represent the chronological progression of the bursts, with lighter colors corresponding to earlier times.
\label{fig:rxte_hysts_epochs}}
\end{figure}

We analyzed the \textit{RXTE} data during the VHE decline (Epochs 3-6), calculating the hardness ratio between the  5--15 keV and 3--5 keV X-ray bands in 10-minute time bins for each of the decline epochs. We then plotted the hardness ratio
as a function of counts in the combined 3-15 keV band to look for any evidence of hysteresis. This is shown in the top panels of Figure~\ref{fig:rxte_hysts_epochs} with a ``zoom-in" shown in the bottom panels corresponding to noticeable rise/decline states in the X-ray (indicated by the solid/colored bands in Figure \ref{fig:xray-VHE-ilc}). All epochs show considerable evolution of the hardness ratio with flux, with a variety of loops and trends. In particular, the top left panel of  Figure~\ref{fig:rxte_hysts_epochs} with observations from Epoch 4 shows an apparent clockwise hysteresis loop between count rates of 100--120~s$^{-1}$ and hardness ratios of 0.02--0.06. A zoom-in of part of this loop in the bottom left panel of Figure~\ref{fig:rxte_hysts_epochs} (corresponding to a burst-like feature in the light curve) shows a hardness ratio increase with increasing count rate, followed by a slight, continued increase in hardness ratio as the count rate begins to decrease, and finally a fairly constant hardness ratio even as the count rate decreases significantly.

%%%% FIGURE 13 %%%%%%%%

\clearpage

\acknowledgements
This research is supported by grants from the U.S. Department of Energy Office of Science, the U.S. National Science Foundation and the Smithsonian Institution, and by NSERC in Canada. This research used resources provided by the Open Science Grid, which is supported by the National Science Foundation and the U.S. Department of Energy's Office of Science, and resources of the National Energy Research Scientific Computing Center (NERSC), a U.S. Department of Energy Office of Science User Facility operated under Contract No. DE-AC02-05CH11231. We acknowledge the excellent work of the technical support staff at the Fred Lawrence Whipple Observatory and at the collaborating institutions in the construction and operation of the instrument.

The MAGIC collaboration would like to thank the Instituto de Astrof\'{\i}sica de Canarias for the excellent working conditions at the Observatorio del Roque de los Muchachos in La Palma. The financial support of the German BMBF and MPG, the Italian INFN and INAF, the Swiss National Fund SNF, the ERDF under the Spanish MINECO (FPA2015-69818-P, FPA2012-36668, FPA2015-68378-P, FPA2015-69210-C6-2-R, FPA2015-69210-C6-4-R, FPA2015-69210-C6-6-R, AYA2015-71042-P, AYA2016-76012-C3-1-P, ESP2015-71662-C2-2-P, FPA2017‐90566‐REDC), the Indian Department of Atomic Energy, the Japanese JSPS and MEXT and the Bulgarian Ministry of Education and Science, National RI Roadmap Project DO1-153/28.08.2018 is gratefully acknowledged. This work was also supported by the Spanish Centro de Excelencia ``Severo Ochoa'' SEV-2016-0588 and SEV-2015-0548, and Unidad de Excelencia ``Mar\'{\i}a de Maeztu'' MDM-2014-0369, by the Croatian Science Foundation (HrZZ) Project IP-2016-06-9782 and the University of Rijeka Project 13.12.1.3.02, by the DFG Collaborative Research Centers SFB823/C4 and SFB876/C3, the Polish National Research Centre grant UMO-2016/22/M/ST9/00382 and by the Brazilian MCTIC, CNPq and FAPERJ.

This publication makes use of data obtained at Mets\"{a}hovi Radio Observatory, operated by Aalto University, Finland, and also the OVRO 40-m monitoring program, which is supported in part by NASA grants NNX08AW31G, NNX11A043G and NNX14AQ89G, and NSF grants AST-0808050 and AST-1109911. The UMRAO data was obtained through NSF grant AST 0607523 and NASA Fermi GI award NNX09AU16G. St.Petersburg University team acknowledges support from Russian RFBR foundation, grant 12-02-00452. The Abastumani Observatory team acknowledges financial support by the Shota Rustaveli National Science Foundation through project FR/577/6-320/13. The Steward Observatory data were obtained under the \emph{Fermi} Guest Investigator Program grant NNX09AU10G. The research at Boston University is supported by NASA grant 80NSSC17K0649 and NSF grant AST-1615796.

\clearpage
%%%%%%%%%%%%%%%%%%%%%%%%%%%%%%%%%%%%%%%%%%%%%%%%%%%%%%%%%%%%%%%%%%%%%%
%%%%%%%%%%%%%%%%%%%%%%%%%%%%%%%%%%%%%%%%%%%%%%%%%%%%%%%%%%%%%%%%%%%%%%
%%%%%%%%%%%%%%%%%%%%%%%%%%%%%%%%%%%%%%%%%%%%%%%%%%%%%%%%%%%%%%%%%%%%%%
\bibliography{Mrk421-2010TextLightCurves}

\end{document}